\documentclass[11pt]{book}
\usepackage{Wiley-AuthoringTemplate}
\usepackage[authoryear]{natbib}

\usepackage{glossaries}
\usepackage{subfigure}
\usepackage{amsmath,amssymb,amsfonts}
\usepackage{graphicx}
\usepackage{textcomp}
\usepackage{wrapfig}
\usepackage{mathtools}
\makeindex
\newacronym{APAS}{APAS}{Almost Perfect auto-correlation Sequences}
\newacronym{MPS}{MPS}{Minimum Peak Sidelobe} 
\newacronym{PMCW}{PMCW}{Phase Modulated Continuous Wave}
\newacronym{MF}{MF}{Merit Factor}
\newacronym{SNR}{SNR}{Signal to Noise Ratio}
\newacronym{INR}{INR}{Interference to Noise Ratio}
\newacronym{SINR}{SINR}{Signal to Interference plus Noise Ratio}
\newacronym{AF}{AF}{Ambiguity Function}
\newacronym{MIMO}{MIMO}{Multiple Input Multiple Output}
\newacronym{SISO}{SISO}{Single Input Single Output}
\newacronym{CD}{CD}{Coordinate Descent}
\newacronym{BCD}{BCD}{Block Coordinate Descent}
\newacronym{GD}{GD}{Gradient Descent}
\newacronym{MM}{MM}{Majorization-Minimization}
\newacronym{FMCW}{FMCW}{Frequency Modulated Continuous Wave}
\newacronym{CDM}{CDM}{Code Division Multiplexing}
\newacronym{DFT}{DFT}{Discrete Fourier Transform}
\newacronym{FFT}{FFT}{Fast Fourier Transform}
\newacronym{MVDR}{MVDR}{Minimum Variance Distortionless Response}
\newacronym{MBI}{MBI}{Maximum Block Improvement}
\newacronym{RFPA}{RFPA}{Radio Frequency Power Amplifier}
\newacronym{BPSK}{BPSK}{Binary Phase Shift Keying}
\newacronym{QPSK}{QPSK}{Quadrature Phase Shift Keying}
\newacronym{ULA}{ULA}{Uniform Linear Array}
\newacronym{DOF}{DOF}{Degrees of Freedom}
\newacronym{PSK}{PSK}{Phase Shift Keying}
\newacronym{PSL}{PSL}{Peak Sidelobe Level}
\newacronym{ISL}{ISL}{Integrated Sidelobe Level}
\newacronym{ISLR}{ISLR}{Integrated Sidelobe Level Ratio}
\newacronym{SILR}{SILR}{Spectral Integrated Level Ratio}
\newacronym{LFM}{LFM}{Linear Frequency Modulation}
\newacronym{HPM}{HPM}{Hybrid Phased MIMO}
\newacronym{MPSK}{MPSK}{$M$-ary Phase Shift Keying}
\newacronym{LPI}{LPI}{Low Probability of Intercept}
\newacronym{RoC}{RoC}{Radar-on-Chip}
\newacronym{RF}{RF}{Radio-Frequency}
\newacronym{PAR}{PAR}{Peak-to-Average Power Ratio}
\newacronym{LTE}{LTE}{Long Term Evolution}
\newacronym{DL}{DL}{Down Link}
\newacronym{UL}{UL}{Up Link}
\newacronym{iid}{i.i.d.}{independent and identically distributed}
\newacronym{BS}{BS}{Base Station}
\newacronym{BSUM}{BSUM}{Block Successive Upper-bound Minimization}
\newacronym{SDR}{SDR}{Software-Defined Radio}
\newacronym{OTA}{OTA}{Over-The-Air}
\newacronym{USRP}{USRP}{Universal Software Radio Peripheral}
\newacronym{ICCL}{ICCL}{Integrated Cross Correlation Level}
\newacronym{ADMM}{ADMM}{Alternating Direction Method of Multipliers}
\newacronym{SDPM}{SDPM}{Spectrum Discrete Phase Modulation}
\newacronym{CW}{CW}{Continuous Wave}
\newacronym{DoA}{DoA}{Direction of Arrival}
\newacronym{MUSIC}{MUSIC}{Multiple Signal Classification}
\newacronym{BiST}{BiST}{Binary Sequences seTs}
\newacronym{PDSCH}{PDSCH}{Physical Downlink Shared Channel}
\newacronym{PDCCH}{PDCCH}{Physical Downlink Control Channel}
\newacronym{MCS}{MCS}{Modulation and Coding Schemes}
\newacronym{GUI}{GUI}{Graphical User Interface}
\newacronym{MI}{MI}{mutual information}
\newacronym{NI}{NI}{National Instruments}
\newacronym{HW}{HW}{hardware}
\newacronym{PAC}{PAC}{Perception/Action Cycle}
\newacronym{RFI}{RFI}{Radio Frequency Interference}
\newacronym{SD}{SD}{Steepest Descent}
\newacronym{CPI}{CPI}{Coherent Processing Interval}
\begin{document}
\chapter[Chapter 12]{Emerging Prototyping Activities in Joint Radar-Communications}
\author*{Bhavani Shankar M. R}
\author{Kumar Vijay Mishra}
\author{Mohammad Alaee-Kerahroodi}

\address{\orgdiv{Interdisciplinary Centre for Security, Reliability and Trust (SnT)}, 
\orgname{University of Luxembourg}, 
\postcode{L-1855}, 
     \city{Luxembourg}, \street{Avenue J. F. Kennedy}, \country{Luxembourg}}%



\address*{Corresponding Author: Bhavani Shankar M. R \email{Bhavani.Shankar@uni.lu}}

\maketitle

\begin{abstract}{Abstract}
The previous chapters have discussed the canvas of joint radar-communications (JRC), highlighting the key approaches of radar-centric, communications-centric and dual-function radar-communications systems. Several signal processing and related aspects enabling these approaches including waveform design, resource allocation, privacy and security, and intelligent surfaces have been elaborated in detail. These topics offer comprehensive theoretical guarantees and algorithms. However, they are largely based on theoretical models. A hardware validation of these techniques would lend credence to the results while enabling their embrace by industry. To this end, this chapter presents some of the prototyping initiatives that address some salient aspects of JRC. We describe some existing prototypes to highlight the challenges in design and performance of JRC. We conclude by presenting some avenues that require prototyping support in the future.
\end{abstract}
%
%
\section{ Motivation}
In recent years, radars that share their spectrum with wireless communications have gained significant research interest \cite{dokhanchi2019mmwave,duggal2020doppler,elbir2021terahertz}. Such a joint radar-communications (JRC) system is being considered for its potential advantages including efficient spectrum and hardware utilization and enhanced situational awareness in applications not limited to automotive. The JRC paradigm is also being considered in the sixth-generation (6G) deployment under  {\em integrated sensing and communications}. While lower frequencies offer fertile ground for the implementation of JRC systems because of spectrum congestion and hardware (HW) availability \cite{paul2016survey}, the wide bandwidth millimeter-wave (mmWave) bands are interesting from the perspective of the emerging high resolution and high data rate applications \cite{mishra2019toward}. 

From a system design objective, the JRC techniques need to devise ways toward improved integration of both sensing and communications systems, wherein finite radio resources must meet the demands of co-existence \cite{wu2022resource} or co-design \cite{liu2020co} regimes. This multi-function approach seeks to develop architectures and algorithms including signaling strategies, receiver processing and side-information  to support their simultaneous operation while meeting appropriately chosen design metrics. Many of these aspects have been addressed in the literature and have been concisely captured in the earlier chapters. These seminal works, as in many of the classical signal processing and communications (SPCOM)  methodologies, have largely taken a model-based approach to the optimization and evaluation of the proposed methodologies for waveforms and receivers. This approach allows for tractability of analysis and eases design. Typical assumptions include the Gaussian modeling of signals, noise and channels, linearity in responses, device uncertainties, isotropic scattering models, absence of clutter, instantaneous availability of information among others. In many a circumstance, these assumptions do not hold, thereby requiring a re-evaluation of the efficiency of these techniques. As a case in point, power amplifiers are not essentially linear, certain types of device noise like shot/impulse noise are not Gaussian and so is signal-correlated quantization noise in low-bit systems \cite{kumari2020low,pace2000advanced,tsui2004digital}. Software simulations incorporating these perturbations do offer certain guarantees, but these simulations are based on models for perturbations, which, again, may suffer from modeling inaccuracies. At mmWave, special considerations are required for designing wideband receivers \cite{mishra2012frequency,tsui2010special,daniels2017forward}.

Towards stepping into the next stage of technological maturity, it becomes essential to validate the JRC concepts, even in controlled, but representative scenarios. This warrants {\em prototyping}, an essential aspect of disseminating SPCOM research of late \cite{mishra2019cognitive}. 
%
It has been well-understood that prototyping enables development teams to explore concepts, understand technical challenges, specify product requirements, and reduce uncertainty. 
%
In this context,  it is desired that the HW prototyping of JRC systems achieves the following objectives \cite{Thesis_HW,RTUN}:
\begin{itemize}
    \item Ascertain the feasibility of JRC algorithmic implementation on HW platform 
    \item Identify and mitigate errors and incorporate missing functionality at an early stage; typical errors would be in synchronization, gain control, alignment and functionalities beyond the physical layer may be required (e.g., channel feedback protocol)
    \item Incorporate the device and scenario modelling errors implicitly
    \item  Gain confidence in the achieved JRC performance and pitch it for consideration in large scale demonstrations
\end{itemize}
\section{Prototyping : General Principles and Categorization}
The following principles are generally considered toward the development of a HW prototype  \cite{RTUN},
\begin{enumerate}
    \item {\bf Identification of the prototyping  objectives:} The first step is a clear enumeration of the demonstration objectives from the prototyping. Clearly, considering  all the aspects of the invention is desirable; however,  several factors including the resources and budget impact this choice. Another aspect is the potential use of this prototype in valorization. Such an objective listing provides an overview of the device and platform requirements for demonstration.
    \item {\bf Prototype High-level Design:}  Based on the objectives, the components and the experimental set-up are identified and a high level design is carried out to ascertain feasibility. This identifies and eliminates system level uncertainties and consolidates the interfaces among components and software.
    \item{\bf Prototype Realization and Evaluation:} Subsequently, the prototype is developed and standard unit and integration tests performed. An elaborate testing is undertaken in the chosen scenario, collecting necessary metrics and performing statistical analysis. These are then compared to system level software simulators for conformance.
    \item{\bf Feedback and Refinement : } The evaluation, provides valuable feedback and data for refining  the prototype;  significant deviations from the software simulator are of particular importance.   This stage also provides the researchers the confidence to pursue further towards enhancing creating a minimum viable product (MVP) and/ or enhance the Technology Readiness Level (TRL) of their invention.
\end{enumerate}
%
%
%
Several categorizations of prototypes exist when considering the product development cycles; one such categorization is as follows \footnote{Based on {\em Four Types of Prototypes and Which is Right for You?},
Voler Systems,  https://www.volersystems.com/blog/design-tips/4-types-prototypes-right, Last Accessed 02 November 2021} :
\begin{enumerate}
    \item Proof of concept prototypes: This is typically the first attempt at prototyping with the purpose of proving the technical feasibility of the research idea.
    \item Looks-like prototypes : This mock-up depicts the {\em visual aspects} of the final product but does not contain working parts. It enables an understanding of the  physical appearance of the product.
    \item Functional or Works-Like Prototypes : These prototypes are used for testing the key functionalities without the looks of the final product.  It complements the Look-like prototypes.
    \item Engineering Prototypes : This would be the final development prior to full production; it is undertaken done to ensure quality and manufacturability.
\end{enumerate}
Table \ref{tab:Table_desc}  summarizes the different design aspects for prototyping \footnote{ Adapted from {\em Types of Hardware Prototypes: Requirements and Validation}, Cadence PCB Design and Analysis Blog,  https://resources.pcb.cadence.com/blog/2019-types-of-hardware-prototypes-requirements-and-validation, last accessed November 2, 2021}.
    \begin{table}
\caption{Design Aspects for Prototyping.\label{tab:Table_desc}}{%
\begin{tabular}{|c|c|}
\toprule
 {\bf Prototyping Aspect} & {\bf Influencing parameters}  \\
\midrule
Regulatory & Spectrum Allocation, Transmission power \\ \hline
Hardware &  Requirements including reconfigurability, Cost, Interfaces,  \\ \hline
Form factor &  Portability, availability of ICs \\ \hline
Power & Battery operated (for portable), socket powered \\ \hline
Software & Operating System, Embedded Firmware, APIs \\ \hline
Durability & Handled at different Temperatures, Humidity, Personnel \\ \hline
\botrule
\end{tabular}}{\footnotetext[]}
\end{table}
The prototyping relevant to this chapter is the  {\bf Proof of Concept.}  The prototypes described in the sequel use existing materials, parts and components to prove that the idea of JRC is feasible under the controlled environment of a lab or a small scale deployment. Within the proof of concept, several possibilities exist:
\begin{itemize}
    \item {\bf Full-fledged HW testbed with over-the-air (OTA) testing:} In this mode, the data acquisition, storage and processing is undertaken in HW, developed on an appropriate platform and embedded software. It uses OTA transmissions to achieve the goal. Such a development enables real-time demonstration using stand alone HW, thereby indicating an advanced prototyping. The HW is typically enabled by widely-available software defined radio (SDR) platforms offering varying degrees of programmability on the included FPGA. 
    \item {\bf Full-fledged HW testbed without over-the-air (OTA) testing:} In many situations, the OTA transmissions are replaced by wired connections with the impact of channel emulated on the FPGA. These situations arise when license to transmit in a particular band is restricted, unavailable or pending (Eg, in Luxembourg, 2400 - 2483.5 MHz Band is reserved only for Short Range Devices with a maximum EIRP of 10mW, which precludes from outdoor applications), avoid radio-frequency (RF) interference to other set-ups or due to lack of capability in the device and/ or experimental set-up to achieve desired goals when performing OTA transmissions. A case in point regarding the last situation is the use of low-cost sub-6GHz SDRs with 100-150 MHz bandwidth for radar applications. These require tens of metres between the antennas and the target to demonstrate any meaningful result on range resolution. This requirement is clearly not met in limited areas with OTA transmissions. 
    \item {\bf Hardware in the loop: } In situations where the HW lacks the needed computation power, in early stages of prototyping or in the use of commercial-off-the-shelf (COTS) components, it would be necessary to include processing on a powerful host, typically a personal computer. While the host computation in software enables easier programming in widely-used Matlab \footnote{https://www.mathworks.com}, LabVIEW \footnote{https://www.ni.com/en-us/shop/labview.htm}, C/ C++ etc, the interface between the HW and SW may offer bottlenecks based on the rate of acquisition. In sub 6GHz based SDR platforms, the bandwidths are the order of 100-200 MHz, thereby allowing seamless transfer of data from the HW platform for host processing resulting in a real-time application. On the other hand, in COTS solutions like TI AWR, Infenion RXS*, the bandwidths are in the range of a 1-4 GHz; using  these COTS for developing applications on the module itself other than those provided by the manufacturer tends to be rather difficult. Further, porting such high speed data onto a computer is also difficult thereby rendering  real-time applications infeasible. 
\end{itemize}
%
%
 %
\section{JRC Prototypes:Typical Features and Functionalities}
%
%
Communication prototypes have been considered for long and have got a fillip from the standardization activity involving academia and research alike. Similarly, radar prototyping in academia has gathered pace, of late, with the development of low-cost COTS sensors. These prototypes demonstrate a number of features and functionalities that are characterized below.
\subsection{Operational Layer}
The communication protocol stack is defined by the seven  Open Systems Interconnection (OSI) layers and HW oriented  prototypes focus on the lowest  Physical layer (PHY) and  the one above, the Medium Access Control (MAC) layers. At the PHY layer, the prototypes implement a  chosen air-interface (e.g, LTE/ A, 5G-NR etc) including the associated frame format, baseband processing and associated RF operations. Some prototypes focus only on waveform (e.g., OFDM, Single Carrier), while keeping the framing to a minimum; prototypes dealing with end-to-end PHY performance include complete framing. On the other hand, prototypes  with integrated MAC layer implement the associated functionalities like packet scheduling, ordering, transmissions/ re-transmissions, handshaking, backoff etc. Finally, the higher layers are typically software oriented and integrated using known interfaces. As a case in point,  open source (GNU GPLv2) discrete-event network simulators are available in C++ for Network and higher layers, e.g., NS3 Simulator, https://www.nsnam.org. 
%
\subsection{Operational Frequencies}
Despite the maturity of resource reuse, the limited bandwidth in lower frequencies, particularly below 6 GHz, precludes the complete exploitation of the 5G potential as well as meeting the 5G requirements. This also reduces the range resolution in radars, thereby precluding emerging applications requiring cm-length accuracy. Further, the equipment at lower frequencies are bulky as well. These call for the use of mmWave spectrum due to the large available bandwidths and high degree of miniaturization. However, a significant communication prototyping activity has focussed on the lower frequency bands due to availability of low-cost hardware platforms. On the other hand, academic radar prototypes in lower frequencies are hard to find for reasons mentioned above. Further, the communication prototypes are in debt at mmWave frequencies. Infact, the implementation in such frequencies is significantly complicated compared to sub-6 GHz  systems due to the following \cite{COM_Mag}
\begin{itemize}
\item Use of highly integrated manufacturing technologies and large-scale phased arrays that contain many radio frequency (RF) analog components. 
\item  Expensive mmWave HW components 
\item  mmWave analog components suffer greatly from different HW imperfections and constraints that severely affect system performance, including phase noise, power amplifier nonlinearities, inphase and quadrature (IQ) imbalance.
\end{itemize}
Nonetheless,  significant research and investment from academia and industry is being made to create flexible and scalable HW solutions \cite{Malik}. This coupled with advances in SDR platforms, novel HW/ SW interfaces and HW abstraction as well as easier programming of HW devices have motivated the design of 5G HW prototypes in mmWave bands. Several issues arise when implementing {\em flexible and simple} HW prototypes in mmWave \cite{Malik}; these impact, RF front ends, data conversion, storage, processing and distribution, processing algorithms, real-time operation among others.  On a similar note, mmWave COTS modules from {\bf TI, Infenion} offering starter-kits with the flexibility of on-chip processing or raw data acquisition, has enhanced the uptake of radar prototyping, mainly for indoor scenarios.   
%
%
%
%
\subsection{MIMO Single User architectures}
Since two decades, the use of multiple antennas in wireless communications has been the norm. The simpler prototypes demonstrating MIMO technology do so for the case of point-to-point link wherein signals received on the multiple antennas can be jointly processed. In this context, the classical VBLAST from Bell labs offered the first prototype operating at a carrier frequency of 1.9 GHz, and a symbol rate of 24.3 ksymbols/sec, in a bandwidth of 30 kHz. The prototype exploits an antenna array, with results reported for 8 transmit and 12 receive antennas \cite{VBLAST}.  Subsequently, significant prototyping efforts focussed on single user or point-point PHY links emulating communication between a transmitter and a particular receiver. These designs assumed existence of orthogonal allocation of resources (e.g., time, bandwidth) across different users (receivers). This led to the sufficiency of prototypes emulating a single-user link on Additive White Gaussian Noise (AWGN) channels to be representative of the system performance.

\subsection{MIMO  Multi-User (MU-MIMO) architectures}
On the other hand, with an increasing demand for data inducing a scarcity of resources, the reuse of resources to serve multiple users offers an attractive alternative, albeit, at the cost of interference among them.  In order to manage the interference, it is essential for the transmitter to know the channel conditions to the served users, necessitating a feedback from different terminals. In addition to requiring multiple receiver HW prototypes demodulating and decoding data corresponding to different users, the multi-user prototype should also support channels for information flow from each of the user to the transmitter. This return link implementation requires protocol support to manage  and  synchronize feedback.  It should be noted that such an implementation for single user channels is simpler due to a single return link. 
%
 
MU-MIMO has become an essential feature of communication prototypes emulating functionalities of 3G standard and beyond. The functionalities implemented in these prototypes includes multiple precoding techniques, power allocation and user scheduling at the transmitter, receive beamforming/ equalization, synchronization, channel and transmission quality feedback.
\subsection{Massive MIMO}
A recent trend is to employ a large number of antennas at the base-station (BS) to exploit spatial diversity combat small scale fading. This Massive MIMO technology has attracted significant interest due to the potential gains offered by theory. However, practical implementation is in debt, partly due to  many critical practical issues that need to be addressed. An increased number of interconnections, much larger processing loads, and enhanced system complexity and cost have been listed as some of the challenges that need be addressed \cite{Ove}. 

Massive MIMO prototyping differs from the conventional approach both in analog as well as digital processing. With an increase in the number of antennas, the analog components feeding those antennas also increase; to ensure a cost effective solution, low-cost components would be used, which tend to be non-ideal. These need to be considered during prototyping to ensure adequate reflection of actual performance. These include the antenna and analog front end imperfections, lack of non-reciprocity and calibration for TDD schemes and computational power.  From a digital processing, a typical choice would include  SDRs and additional computational resources to ensure flexibility and real-time operations. To exploit the gains offered by large number of antennas, sample level time synchronization and phase alignment among the different streams becomes crucial. Another important aspect is large amounts of data that need to be acquired, stored and moved around different processing units; the underlying HW platform should be able to handle this efficiently. Finally, the baseband processing should implement transmission and reception techniques that could be rather involved. 

%
\subsection{Phased Array and MIMO Radar}
Phased-array radar systems have been considered since long originating from the military applications towards  detecting and tracking objects small objects at significantly large distances by concentrating power in a sharp beam that is electronically steered at high rates \cite{PA_hist}. With the proliferation of radar in civilian applications including automotive and indoor sensing, there has been the inclusion of additional constraints including (i) wide-angle operation to have an increased coverage, (ii) wide-band processing leveraging on the high available bandwidth in contrast to traditional narrow-band systems and (iii) Restrictions on size, weight, cost and power. Several avenues have been considered towards addressing one or many of these problems. To reduce the the cost of the RF components for beamforming and signal distribution, RF Systems on a Chip (RFSoC) have been considered to minimize cost and offer implementation of SP algorithms in real-time \cite{RFSOC}. A prototype RFSoC with 16 as a software defined receiver and waveform generator along with real-time adaptive beamforming in an S-band phased array radar. The focus of the research has been to demonstrate the potential improvement for size weight and power (SWaP), real-time signal processing capacity and advanced design processes for rapid algorithm implementation. The pre-production RFSoC prototype demonstrates the potential value of RFSoC and rapid algorithm development for next generation radar systems.

In this context, the current developments have focussed on the design and development of large-scale phased array antennas with tunable phase shifter

%
\subsection{Summary}
With the increased interest, support and availability of HW, prototyping, particularly in research has seen an increased uptake in the recent years. In the case of wireless communications, Table \ref{tab:Comm_Ex} presents some representative examples of different prototypes to highlight the changes in HW and SW. The current trend has been on the 5G prototyping with several initiatives world-wide on developing testbeds and are numerous to list here. An useful resource could be the IEEE Future Networks Initiative (FNI) which curates a directory of available 5G and beyond networking testbeds for use by both academic and industry research groups \footnote{https://futurenetworks.ieee.org/testbeds}.  
   \begin{table}[h]
\caption{Examples of Wireless Communication Prototypes.\label{tab:Comm_Ex}}{%
\begin{tabular}{|c|c|c|c|}
\toprule
 {\bf Prototype} & {\bf WARP} & {\bf SAMURAI} & {\bf KU Leuven} \\
\midrule
         {\bf Reference} & \cite{warpProject} & \cite{SAMURAI} & \cite{KUL}\\  
          {\bf Protocol} & 802.11, OFDM & LTE Rel 8 TM6. & LTE-like (TDD)\\   
          {\bf Architecture} & SISO-SU &  MU-MIMO & Massive MIMO\\  
          {\bf Transmit Antennas} & 1 &  4 (ExpressMIMO) & 68  \\  
          {\bf Frequency} & 2.4/ 5 GHz & 250 MHz‐3.8 GHz \footnote{https://cordis.europa.eu/docs/projects/cnect/6/257626/080/deliverables/001-ACROPOLISD52v10.pdf}  & 400 MHz - 4.4 GHz\\  
          {\bf Layers} & PHY, MAC & PHY, MAC & PHY \\  
          {\bf HW} &  Virtex-6   & Virtex 5 LX330 & Kintex 7\\  
             &  LX240T FPGA  &  Virtex 5 LX110T &  \\  
           {\bf SW} &  C &  C & LabView \\  \hline
\botrule
\end{tabular}}{\footnotetext[]}
\end{table}

On a similar trend, there have been a number of radar test beds in research institutions implementing various radar waveforms (FMCW/ PMCW/ Pulsed/ CW) and undertaking different radar tasks. Unlike the wireless communications, lack of a standardized approach results in different prototypes using different configurations. In this context, rather than an elaborate listing, Table \ref{tab1-2} collects the key components from modern radar prototyping.
    \begin{table}
\caption{Typical Components of Radar  Prototyping.\label{tab1-2}}{%
\begin{tabular}{|l|l|l|l|}
\toprule
 {\bf HW } &  {\bf Enabling} & {\bf Potential} &{\bf Possible} \\
 {\bf  Platform} &  {\bf  Software} & {\bf Functionality} &{\bf Waveforms} \\ \midrule
USRP NI-29xx &  LabView NXG
 & MIMO Radar & FMCW\\
sub 6GHz&  Matlab & Cognitive radar & PMCW \\
National&  Qt C++ & Coexistence & CW \\
Instruments & Python & JRC & Custom\\ \hline
COTS  Modules   & MATLAB 
& Distributed sensing & FMCW\\
(AWR , IWR,  …)  & Python & Imaging & \\
24, 60, 77-81 GHz & Code composer studio  & Tracking  & \\
 Texas Inst. &  C++ &Interference analysis & \\ \hline
FPGA boards &  VHDL &  Cognitive radar & FMCW\\ 
mmWave & ISE/ Vivado & Customized MIMO radar & PMCW\\
Xilinx & Python & DoA Finding & CW\\
 & HLS/C++ & Customized SP units & Custom \\ \hline
\botrule
\end{tabular}}{\footnotetext[]}
\end{table}
\section{JRC Prototyping}
The previous sections provided an overview and some examples of prototyping in wireless communications and radar. This section, builds on the previous section and brings out additional aspects needed for JRC prototyping.
%
The architecture of the JRC prototype depends on the topology considered. As discussed in the previous chapters, the two predominant topologies are $-$ Co-existence and Co-design exist with regards to the transmission. On the other hand, with regards to the radar reception, the receiver can be co-located with the transmitter (monostatic) or separated (bistatic). In this chapter, we focus on the prevalant monostatic radar implementations. The bistatic operations require additional synchronization and is currently under investigation for the JRC.  The implications of the chosen architecture on the prototyping are detailed below.
\subsection{Co-existence}
\begin{figure*}[tbh]	
    \centering
    \subfigure[Bilateral Mode of Operation] {\label{fig:Bilateral}\includegraphics[width=120mm]{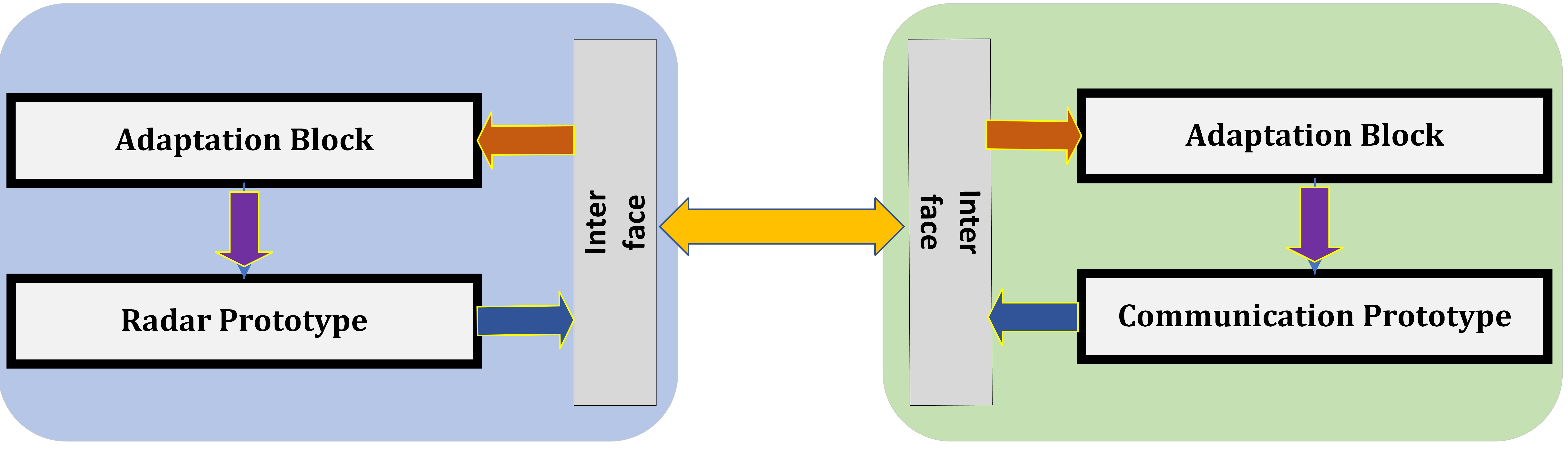}}
   \subfigure[Unilateral Mode of Operation] {\label{fig:Unilateral}\includegraphics[width=120mm]{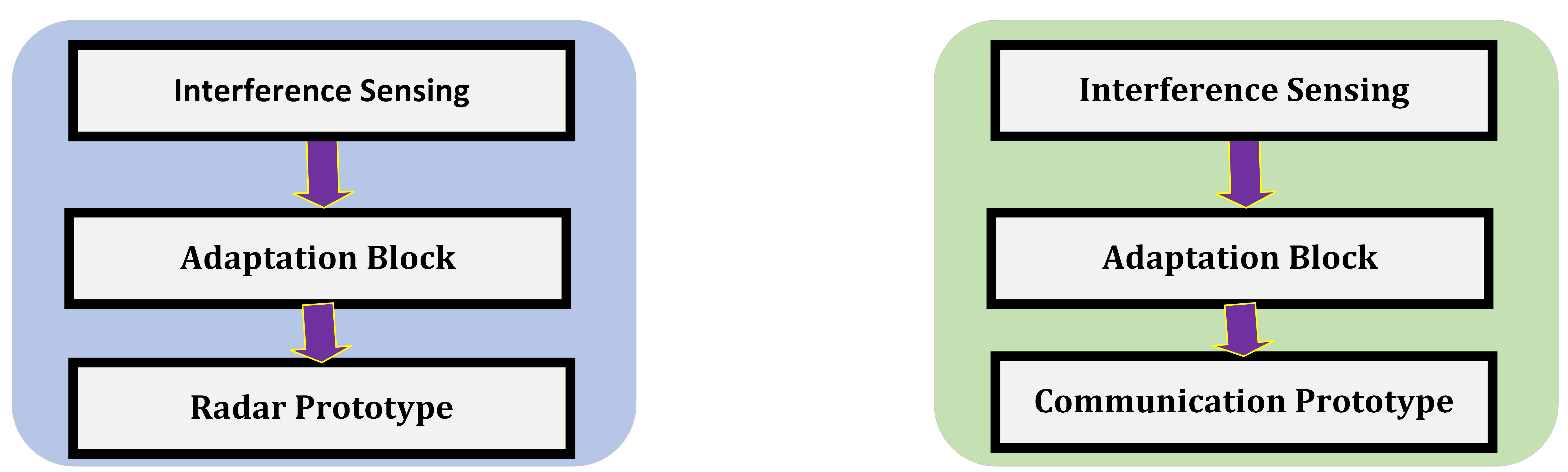}}
    \caption[]{Typical Co-existence Architectures depicting the mode of enforcing co-existence.}\label{fig:Coexistence}
\end{figure*}

In this topology, the radar and communication portions are implemented as separate functionalities as shown in Figure \ref{fig:Coexistence}. To foster co-existence, the following cognitive  paradigms can be considered.
\begin{enumerate}
    \item {\bf Bilateral Operation:} This operation is depicted in Figure \ref{fig:Bilateral}. Here both the systems adapt their transmission to minimize mutual interference. To enable this, an interface link between the two systems enabling high level information exchange (uni/bi directional) is needed and an adaptation module that acts on the information to enable co-existence.
    \item {\bf Unilateral Operation:} This operation is depicted in Figure \ref{fig:Unilateral}.  In this setting, a cognitive module operates in either or both of the systems, that senses the onset of interference and adapts the transmission and reception unilaterally, without additional information exchange. This operation arises in the opportunistic use of spectrum.
\end{enumerate}
In this context, the design critically depends on the nature of operation considered above. In particular, the following aspects need to be carefully considered for {\em Bilateral Operation}
\begin{enumerate}
    \item {\bf Information to be exchanged:} The amount of information, its rate of exchange, the sensitivity of the information and the latency play a fundamental part in achieving. Typical information to be exchanged involves spatial, temporal and spectral resource allocation elements like the power, frequency bands, time-slots and direction of transmission. Unless the resources are large in number (e.g., massive MIMO), exchange of these quantities requires limited bits. Further, the rate of update of these quantities depends on the changes in scenario and can be rather infrequent for nearly static or pre-scanned (known trajectory, user behaviour etc) set-ups. 
    \item {\bf Interface Link Design:} Based on the information to be exchanged, a communication link and an underlying protocol needs to be implemented. This would involve appropriate air-interface based on link-budget and appropriate handshaking protocol. Both wireless and wired designs can be considered; while the former is elegant and is representative of the actual set-up, it needs to address additionally the spectral allocation for the interference link. A wired set-up, on the other hand, enables quick testing of the system idea.
    
    From a HW device perspective, the interface link design entails the need to accommodate additional transmission and reception ports at either end along with appropriate RF front-ends. Dedicated digital processing towards link maintenance (e.g., set-up, handshaking) at either end is also needed.
    \item {\bf Adaptation Block:} This digital processing block needs to be implemented at radar and/ or communication prototypes to adapt the transmission and reception. The operation of this block needs to be synchronized with that of the underlying prototypes, e.g., adaptation needs to be carried out at a pre-defined  intervals which needs to be synchronized either with the transmission frame of the communication or the Coherent processing interval of the radar.
\end{enumerate}
With regards to the  {\em Unilateral Operation}, the host system (radar or communication) needs to incorporate a {\em sensing block} that determines the reuse of resources by the incumbent system and an {\em adaptation block} to act on this information. This block could incorporate an energy detector determining the incident energy in spatial, temporal or spectral domains. Such a block need not be aware of the specifics of the transmission from the incumbent system, thereby simplifying information exchange overhead. From a HW perspective, incorporating such a sensing block would require antenna ports and relevant RF blocks to sense incident waves and a digital processing unit to incorporate the sensing algorithm. The output of the sensing algorithm would be used in the adaptation block mentioned earlier.

\subsection{Co-design}
The co-design prototype involves a single transmitter supporting both functionalities; the bulk of the design lies in the digital processing to derive appropriate transmission strategies with the HW (e.g. power, bandwidth) being dimensioned to ensure the quality of service is met. Differently from the co-existence, the {\em co-design} aspect needs to consider the impact of self-interference. Particularly, when the co-design JRC transmitter is emitting a waveform, it could cause significant self-interference saturating the front end impacting not only the radar, but also the  received communication signal in case of bidirectional transmission. Research on exploiting in-band self-interference cancellation using principles from full-duplex operations have been pursued and some prototypes have been considered \cite{Taneli}. An alternative, but less elegant way, is to isolate the transmit and receive functionalities. 

Having presented the general set-up of the JRC prototypes, the chapter now delves deeper and discusses an example prototype on JRC co-existence in detail based on unilateral operation. Other prototypes would be discussed in the sequel.
\section{Co-existence JRC prototype}
In this section, a JRC prototype enabling spectrum sharing between the radar and communication is presented.  The communication system is assumed to be the primary user of the spectrum and the MIMO radar uses these frequencies in an opportunistic fashion. In this context, the radar performs unilateral operations to sense the spectrum use and exploit it with minimal interference to the communication link. To enable this co-existence, following the blocks in Figure \ref{fig:Unilateral}, the radar module designs:
\begin{enumerate}
    \item {\bf Sensing Block} involves a spectrum sensing application where the granularity of the spectrum block is determined by the resource allocation unit of the communication system. 
    \item {\bf Adaptation Block} exploits the output of the spectrum sensing and adapts the MIMO waveform  {\em on-the-fly}. The sensing and adaptation lead to a {\em Cognitive Radar} set-up where the transmission  is adapted based on the environment.
\end{enumerate}
The focus of this section will be on the prototyping aspects. Details on the waveform design aspects can be obtained from \cite{Sensors}.
\subsection{Architecture}\label{ssec:Arch}
\begin{figure*}[tbh]	
    \centering
    \subfigure[Figure A]{\label{fig:Softwarea}\includegraphics[width=0.3\linewidth]{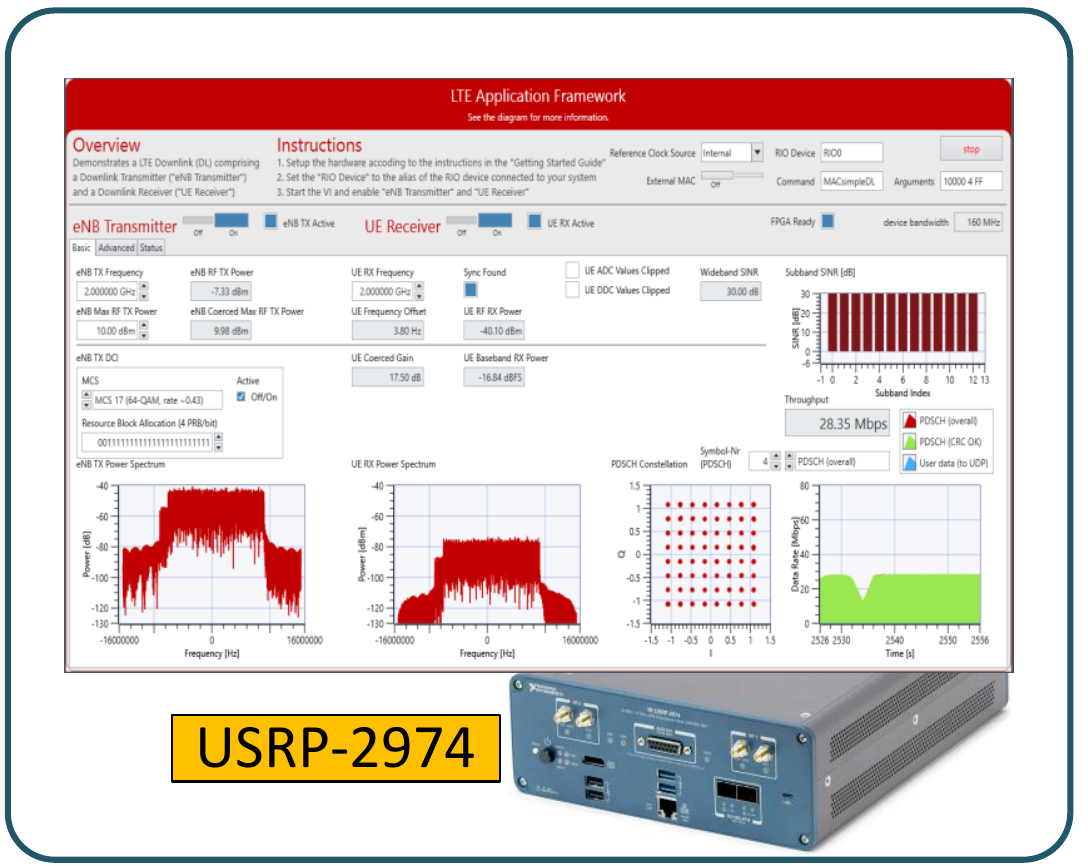}}
    \subfigure[Figure B]{\label{fig:Softwareb}\includegraphics[width=0.3\linewidth]{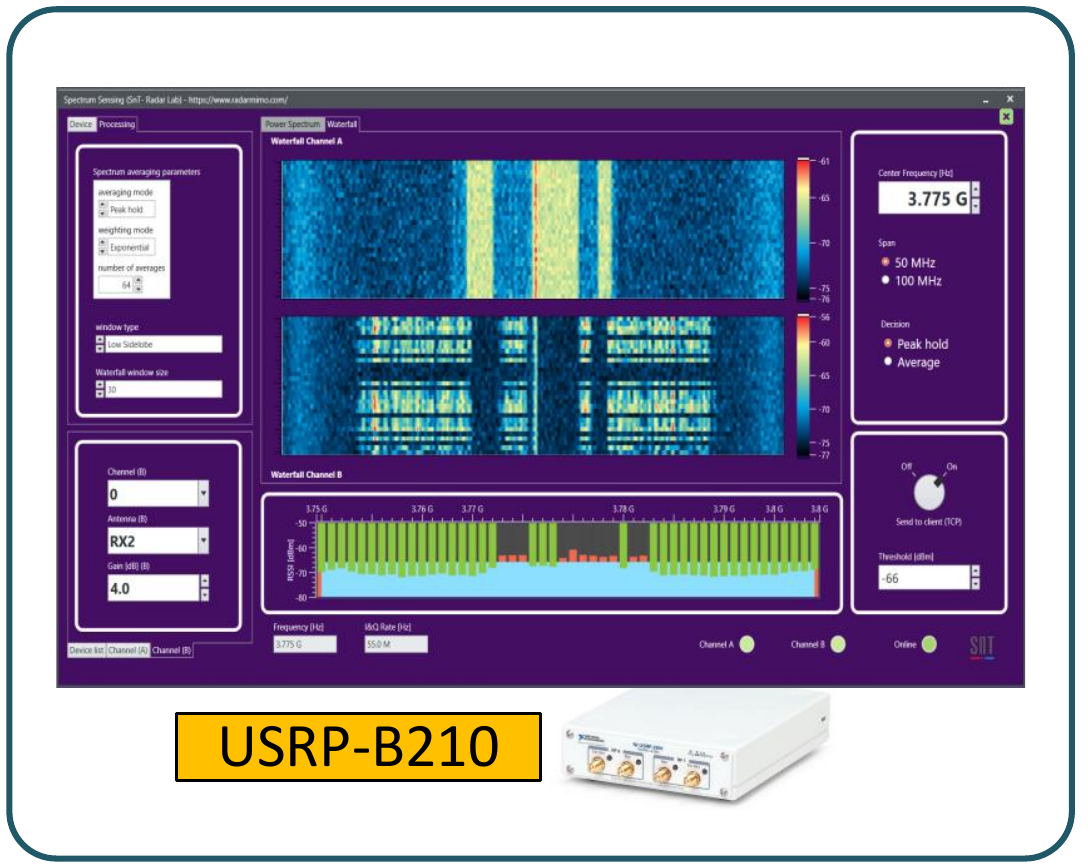}}
    \subfigure[Figure C]{\label{fig:Softwarec}\includegraphics[width=0.3\linewidth]{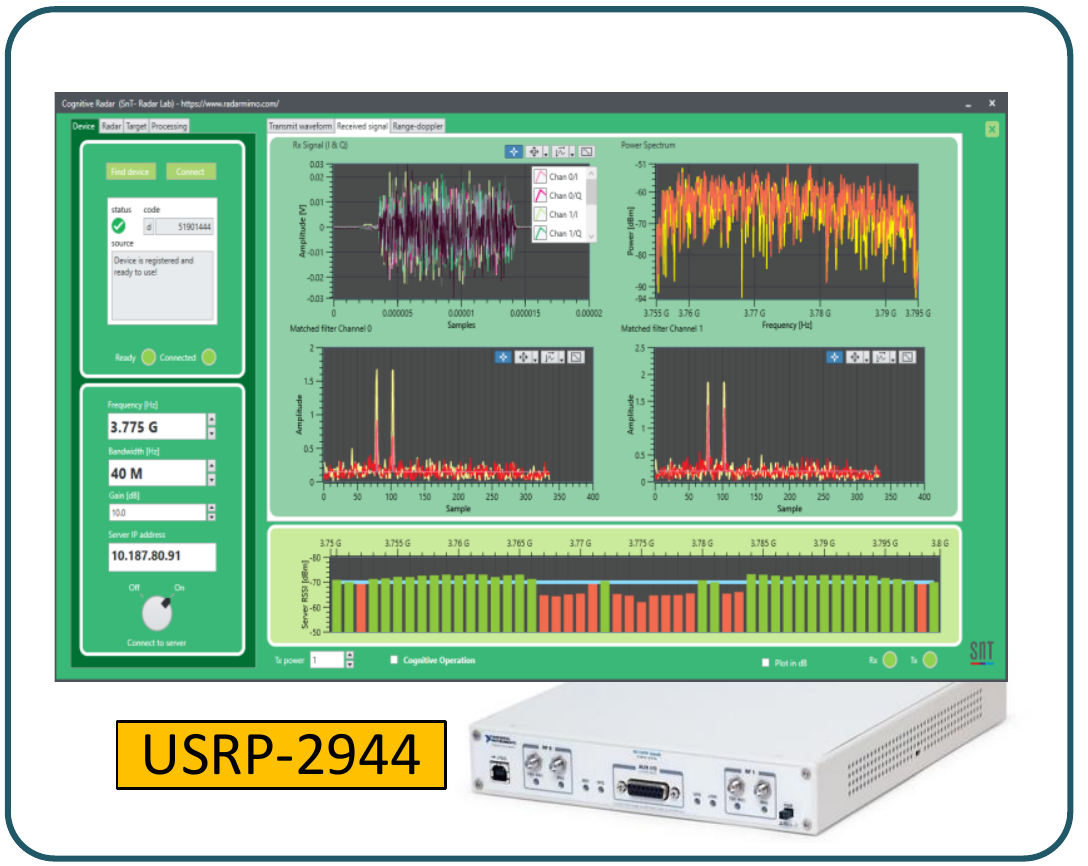}}
    \caption[]{Components or Application frameworks of the prototype: \gls{LTE} application developed by NI, spectrum sensing and cognitive \gls{MIMO} radar applications developed in \cite{Sensors}.}\label{fig:Software}
\end{figure*}

\begin{figure*}
    \centering
    \includegraphics[width=\linewidth]{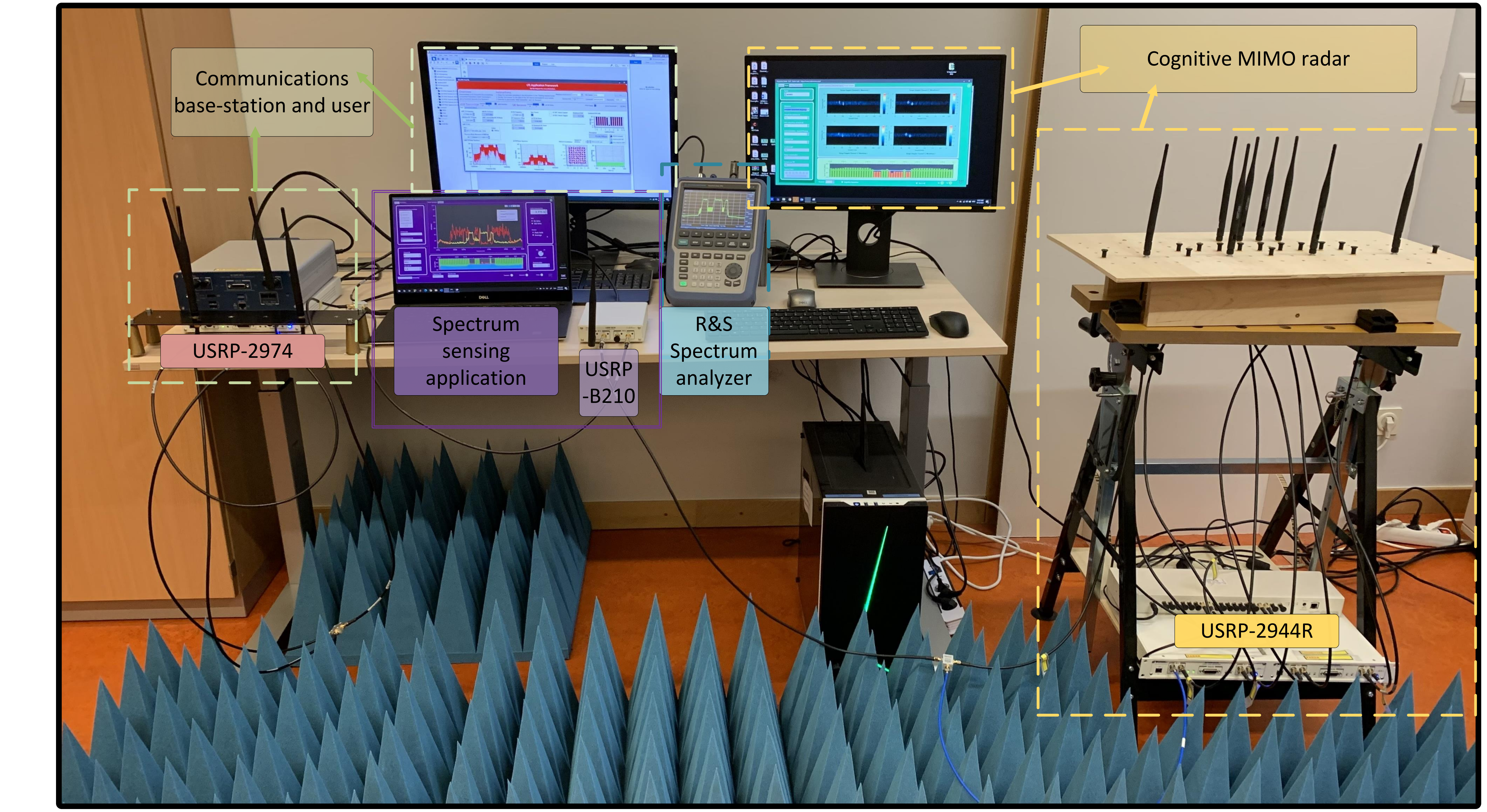}
     \caption{ A photograph  of the proposed coexistence prototype. The photo shows communication \gls{BS} and user, spectrum sensing, and cognitive \gls{MIMO} radar systems.
    }
    \label{fig:Setup}
\end{figure*}

The prototype consists of three components $-$ application frameworks $-$ as depicted in \figurename{~\ref{fig:Software}} along with a photograph of the  set-up in the lab \figurename{~\ref{fig:Setup}} \cite{Sensors}. The three components use \gls{USRP}s for the transmission and reception of the wireless RF signals, the characteristics of which  are presented in \tablename{~\ref{tab:USRPs}}. The transmission is validated using Rohde and Schwarz spectrum analyzer. The components and their associated \gls{HW} are presented below:
\begin{enumerate}
    \item \gls{LTE} Application implemented by National Instruments on \gls{USRP} 2974 and used for \gls{LTE} communications
    \item Spectrum sensing implemented on \gls{USRP} B210 
    \item Cognitive \gls{MIMO} radar implemented on \gls{USRP} 2944R for cognitive \gls{MIMO} 
\end{enumerate} 
\begin{table}
\caption{Hardware characteristics of the proposed prototype. \label{tab:USRPs}}
{%
\begin{tabular}{|c|c|c|}
    \toprule
    \textbf{Parameters}                 & 
    \textbf{B210}       & \textbf{2974/2944R}                                      
    \\ \midrule
    Frequency range                        & $70$ MHz $ - 6$ GHz    & $10$ MHz $ - 6$ GHz                   \\ 
     Max. input power        & $-15$ dBm   & $+10$ dBm              \\ 
    Max. output power                   & $10$ dBm   & $20$ dBm   \\ 
    Noise figure       & $8$ dB  & $5 - 7$ dB    \\
    Bandwidth          & $56$ MHz    & $160$ MHz                          \\ 
    DACs               &  $61.44$ MS/s, $12$ bits  &  $200$ MS/s, $16$ bits     \\ 
    ADCs               &  $61.44$ MS/s, $12$  bits  & $200$ MS/s, $14$  bits    
    \\ 
  \botrule
\end{tabular}}{\footnotetext[]}
\end{table}
\subsection{LTE Application Framework}
The LabVIEW based \gls{LTE} Application Framework (\figurename{~\ref{fig:Softwarea}})  provides a real-time \gls{LTE} physical layer implementation; this add-on software is available in the form of an open and modifiable source-code \cite{LTELabview}. Table \ref{tab:LTE} lists particular 3GPP \gls{LTE} features to which the generated physical layer frame is compliant.
\begin{table}
\caption{\gls{LTE} features of the considered Application Framework.\label{tab:LTE}}
{%
\begin{tabular}{|l|l|}
\toprule
 {\bf Features } &  {\bf Features} \\ \midrule
Closed-loop \gls{OTA} operation & $20$ MHz bandwidth \\  
 channel state and ACK/ NACK feedback &  $\leq 75$ Mbps data-rate\\ 
 $5$-frame structure &  QPSK, $16$-QAM, and $64$-QAM \\  
 \gls{PDSCH}  &  FDD  \\ 
  \gls{PDCCH} &   TDD\\ 
  Channel Estimation & Zero Forcing equalization \\ 
\botrule
\end{tabular}}{\footnotetext[]}
\end{table}
A basic implementation of MAC  enabling packet-based data transmission along with a  framework for rate adaptation is also implemented.

The \gls{NI}-\gls{USRP} 2974 has two independent \gls{RF} chains and the application framework supports single antenna links. In this context, the considered prototype emulates a \gls{SISO} communication link between a \gls{BS} and communications user on the two different \gls{RF} chains of the same \gls{USRP}.
\subsection{Spectrum Sensing Application}
An application based on LabView NXG 3.1  connecting to Ettus \gls{USRP} B210  (\figurename{~\ref{fig:Softwareb}}) has been developed for continuous sensing of the spectrum.  This  application exhibits flexibility and  can update different parameters {\em on-the-fly}, e.g.,  averaging modes, window type, energy detection threshold, and the \gls{USRP} configurations (gain, channel, start frequency, etc.). 
Herein, the center frequency can be set to any arbitrary value in the range $70$ MHz to $6$ GHz, and the  bandwidth spanned can be set to values in the interval $[50 - 100]$ MHz. In this context, it should be noted that \gls{USRP} B210 provides
$56$ MHz of real-time bandwidth by using AD9361 RFIC direct-conversion transceiver. However, efficient implementation enables the developed application to analyze larger bandwidths by sweeping the spectrum. 

The spectrum sensing application determines the specturm occupancy with a resolution of $1$ MHz by computing the energy in the band and a subsequent thresholding \cite{Sensors}. It then  obtains a frequency occupancy chart and transfers this to the cognitive \gls{MIMO} radar application through a network connection (LAN/Wi-Fi). The spectrum sensing is developed on a separate \gls{USRP} and hence, in this context, it can be used as a stand-alone application.
\subsection{MIMO Radar Prototype} 
\begin{figure*}
    \centering
    \includegraphics[width=1.0\linewidth]{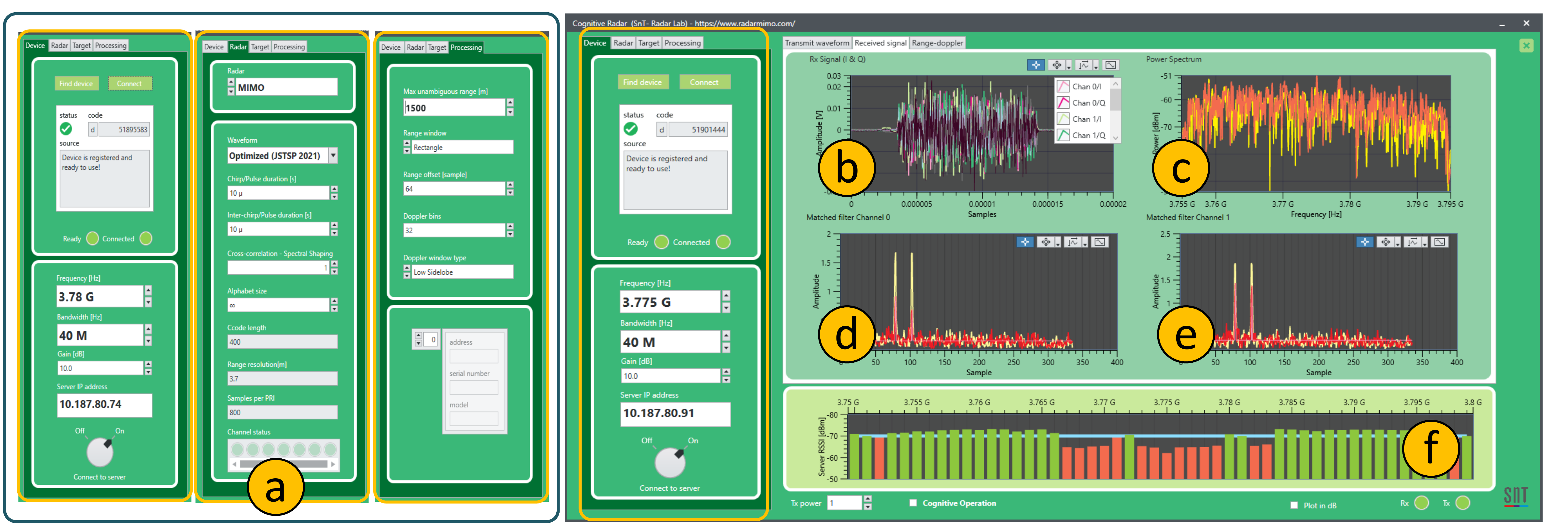}
    \caption[]{Developed cognitive \gls{MIMO} radar application. (a) User interface for setting the parameters of the HW device, radar, and the processing aspects, (b) Baseband I/ Q signals from the two receive channels, (c) Spectrum of the Received signal from the two receive channels, (d) Output of filters matched to the two transmit waveforms on the first receive channel, (e) Output of the  filters matched to two transmit waveforms on the second receive channel, (f)  Side information from the energy detector of the spectrum sensing application. 
    }\label{fig:CognitiveMIMOradar}
\end{figure*}

The developed cognitive \gls{MIMO}  radar application framework is illustrated in \figurename{~\ref{fig:CognitiveMIMOradar}}. The licensed band at $3.78$ GHz with $40$ MHz bandwidth was used for transmission \cite{Sensors}. In this flexible implementation, the parameters related to the radar waveform, processing units, and targets can be updated even during the operational phase to adapt the system to the environment. 
The \gls{MIMO} radar application was developed using LabView NXG 3.1, and  connected to the NI-USRP 2944R comprising $2\times2$ \gls{MIMO} RF transceiver with a  programmable Kintex-7 field programmable gate array (FPGA). 
\tablename{~\ref{tab:Apps}} details the features and flexibility of the developed application. 
\begin{table}[tbh]
\caption{Features of the developed cognitive {MIMO} radar \label{tab:Apps}}{
\begin{tabular}{|c|c|}
\toprule
\textbf{Parameters}                      & 
    \textbf{{MIMO} radar}                         
    \\ \midrule
    \hline
    {Centre Frequency}      & $70$ MHz - $6$ GHz \\ 
    \hline
    {Bandwidth}      & $1-80$ MHz \\ 
    \hline
     {Processing units}         
   & Matched filtering, range-Doppler processing             \\ 
   \hline
   {Window type}         &  Rectangle, Hamming, Blackman, etc.         \\ 
    \hline
    {Averaging mode}         & Coherent integration ({FFT})  \\
    \hline
    {Transmit  waveforms }              & 
    \begin{tabular}[c]{@{}l@{}}
    Random-polyphase, Frank,  Golomb, \\
    Random-Binary, Barker, m-Sequence,\\
    Gold, Kasami, Up-LFM, Down-LFM,\\
    and the optimized sequences
    \end{tabular}    \\ 
   \botrule
\end{tabular}}{\footnotetext[]}
\end{table}

\begin{figure}
    \centering
        \includegraphics[width=0.8\linewidth]{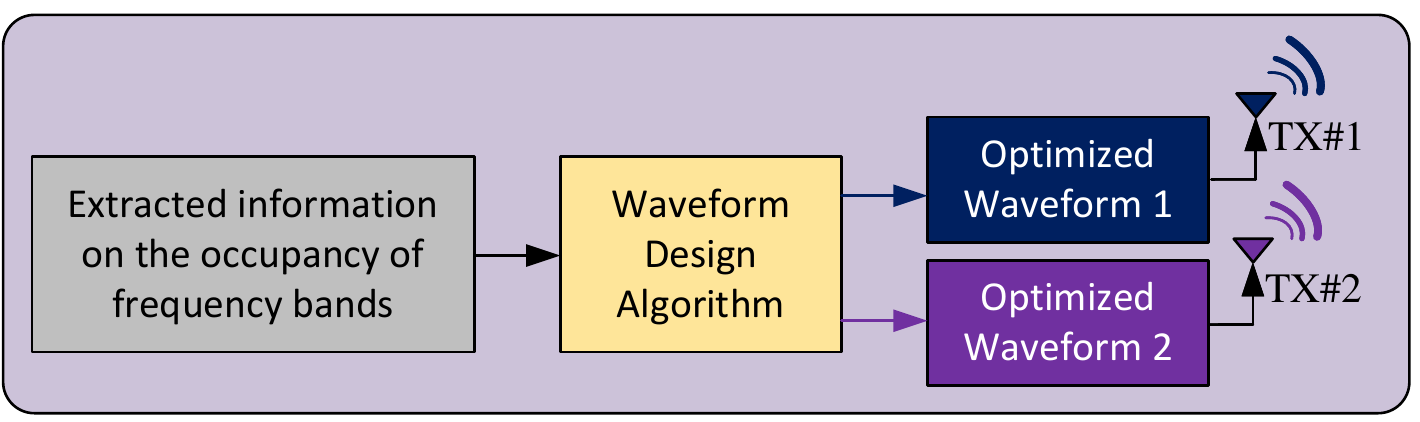}
    \caption[]{Block diagram of the developed cognitive \gls{MIMO} radar transmitter. Information about occupied frequency bands provided  by the spectrum sensing application is used to design 2 waveforms for JRC co-existence \cite{Sensors}.  }
    \label{fig:RadarBlockDiagramTx}
\end{figure}

The transmit  operations of the developed cognitive \gls{MIMO} radar is depicted in \figurename{~\ref{fig:RadarBlockDiagramTx}}. Based on a perusal of the spectrum occupancy list transmitted by the spectrum sensing application over a wireless network, the radar optimizes the transmit waveform, details of which are provided in the sequel. 
\subsection{Waveform Design}
\label{ssec:wave}
Central to the unilateral mode of co-existence is the design of the radar waveform, herein, chosen as the Phase Modulated Continuous Wave (PMCW) type. Herein, the waveform is a sequence of modulated pulses and a certain sequence of pulses form the  pulse code or the waveform to be designed. The duration of the pulse and the length of the pulse code are obtained from system design. Through the flexibility offered in the design of modulation, PMCW waveform offers higher design degrees of freedom compared to other classical approaches like the Frequency Modulated Continuous Wave (FMCW). In this setting, a colocated narrow-band \gls{MIMO} radar system, with $M$ transmit antennas, each transmitting a pulse code or a waveform sequence of length $N$ in the fast-time domain is considered. Denote by matrix ${\mathbf X} \in \mathbb{C}^{M \times N} \triangleq [\mathbf{x}_1^T, \dots, \mathbf{x}_M^T]^T$,  the transmitted set of sequences in baseband, where the vector $\mathbf{x}_m \triangleq [x_{m,1}, \ldots, x_{m,N}]^T \in \mathbb{C}^N$ indicates the $N$ samples of the $m^{th}$ transmitter ($m \in \{1,\dots,M\}$). The aim of the waveform design is two fold:
\begin{enumerate}
    \item To design a  set of transmit sequences having small cross-correlation among each other  to enable their separation at the receiver and enhance the angular resolution property using the concept of MIMO radars. To focus on the \gls{HW} aspects, the detailed modelling of the cross-correlation is omitted with details available in \cite{Sensors}. It suffices to know that there exists a cost function $g_c({\mathbf X})$ whose minimization ensures the required property on cross-correlation.
    \item In addition, the waveforms need to be designed to have certain spectral properties to ensure utilization of unused frequencies while limiting the power transmitted in frequencies that are occupied by the communication signals. This way, the radar can use the full bandwidth, keep the interference low and work without additional knowledge of the location(s) of the communication terminals. As above, the details are available in \cite{Sensors} and it suffices to mention the existence of a cost-function $g_s({\mathbf X})$ that ensures adherence to the desired spectral properties.
\end{enumerate}
Additionally, the PMCW code is drawn from constant modulus constellation to enable efficient power amplification. Define $\Omega_{\infty} = [0,2\pi)$, 
and $\Omega_L = \left\{0, \frac{2\pi}{L}, \dots, \frac{2\pi(L-1)}{L}\right\}$, to a continuous and discrete set of phase values respectively. To ensure constant modulus, one of the constraints defined below is imposed on the design,
\begin{equation}
C_1 \triangleq \{\mathbf{X} \mid x_{m,n} = e^{j\phi_{m,n}}, \phi_{m,n} \in \Omega_{\infty} \}, 
\end{equation}
or 
\begin{equation}
    C_2 \triangleq \{\mathbf{X} \mid x_{m,n} = e^{j\phi_{m,n}}, \phi_{m,n} \in \Omega_L \}.
\end{equation}
To this end, a bi-objective optimization problem considering both the desired objectives, takes the form,
\begin{equation}\label{eq:MOOP-Chapter12}
	\begin{dcases}
	\min_{\mathbf{X}} 	& g_s(\mathbf{X}), g_c(\mathbf{X}) \\
	s.t. 	    & C_1 \ \text{or} \ C_2.
	\end{dcases}
\end{equation}
Such problems are time-consuming to solve; instead, scalarization is a well known technique that converts the bi-objective optimization problem to a single objective problem through a  weighted sum of the objective functions. While scalarization spans the solutions of bi-objective function only in certain settings, such a formulation is nonetheless pursued for the ease of its implementation. The scalarization subsequently leads to the following Pareto-optimization problem,
\begin{equation}\label{eq:sum_weighted-Chap12}
	\mathcal{P}
	\begin{dcases}
	\min_{\mathbf{X}} 	& g(\mathbf{X}) \triangleq \theta g_s(\mathbf{X}) + (1-\theta) g_c(\mathbf{X}) \\
	s.t. 	    & C_1 \ \text{or} \ C_2,
	\end{dcases}
\end{equation}
The coefficient $\theta \in \left[0,1\right]$ is a weight factor effecting trade-off between radar performance and co-existence. The problem in \eqref{eq:sum_weighted-Chap12} is rather difficult to solve and sub-optimal, yet fast and effective, iterative solutions are presented in \cite{Sensors}. These iterative methods naturally support the {\em on-the-fly} waveform adaptation for co-existence whence system parameters can be changed in between iterations without interrupting the flow.
\subsection{Adaptive Receive Processing}
\begin{figure}
    \centering
    \subfigure[Processing unit for range-Doppler creation related to every optimized waveform]{\label{fig:Softwarea}\includegraphics[width=0.8\linewidth]{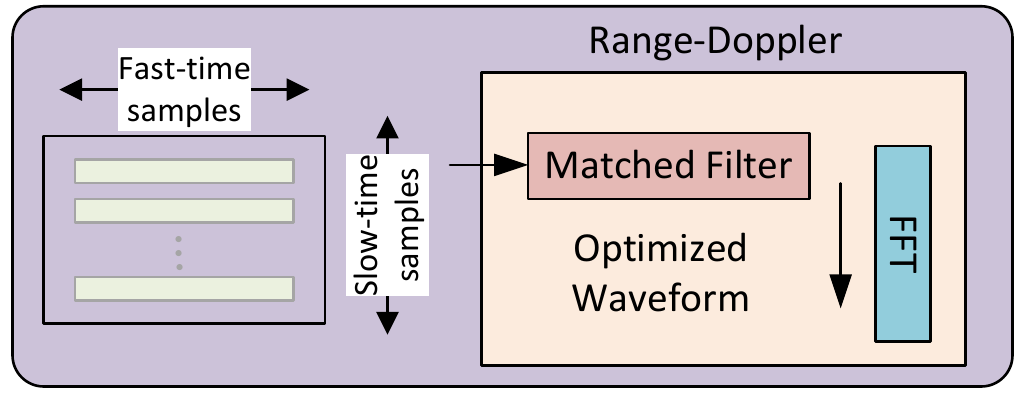}}
    \subfigure[ Block diagram of the  receiver of the developed cognitive \gls{MIMO} radar application.]{\label{fig:Softwarea}\includegraphics[width=0.8\linewidth]{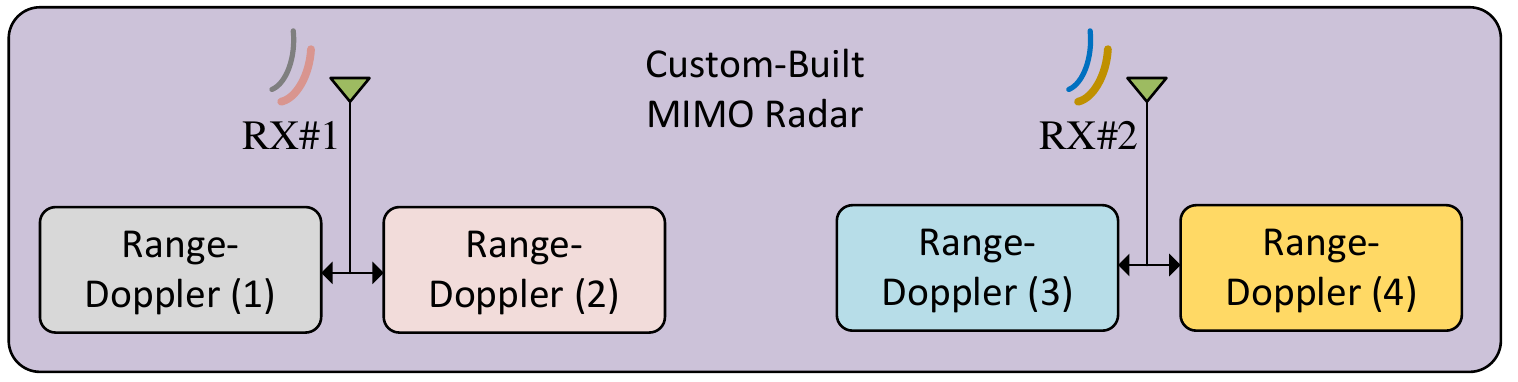}}
    \caption[]{In (a), the matched filter coefficients are updated  for appropriate  filtering in the fast-time domain. Consequently,  the range-Doppler plots are calculated after performing \gls{FFT} in the slow-time dimension. The part related to (a) will be repeated in (b) for every  transmit waveform.    }
    \label{fig:RadarBlockDiagramRx}
\end{figure}

The developed cognitive \gls{MIMO} radar receiver is depicted in Figure \ref{fig:RadarBlockDiagramRx}. 
The sampling at the receiver starts at the onset of  a trigger initiated by transmitter to indicate the onset of transmission.  A classical \gls{MIMO} radar operation is undertaken where, in each receive channel, two filters matched to each of the two transmitted waveforms is implemented using the \gls{FFT} based  technique. Consequently, a total of 4 matched filter outputs are obtained at the receiver and four range-Doppler plots corresponding to the receive channels and transmitting waveforms are obtained by implementing \gls{FFT} in the slow-time dimension.

The receiver processing unit adapts the matched filter as and when the transmit waveform is changed.
\subsection{Experimental Set-up} 
\label{Sec:Experiments}
\begin{figure}
    \centering
        \includegraphics[width=1\linewidth]{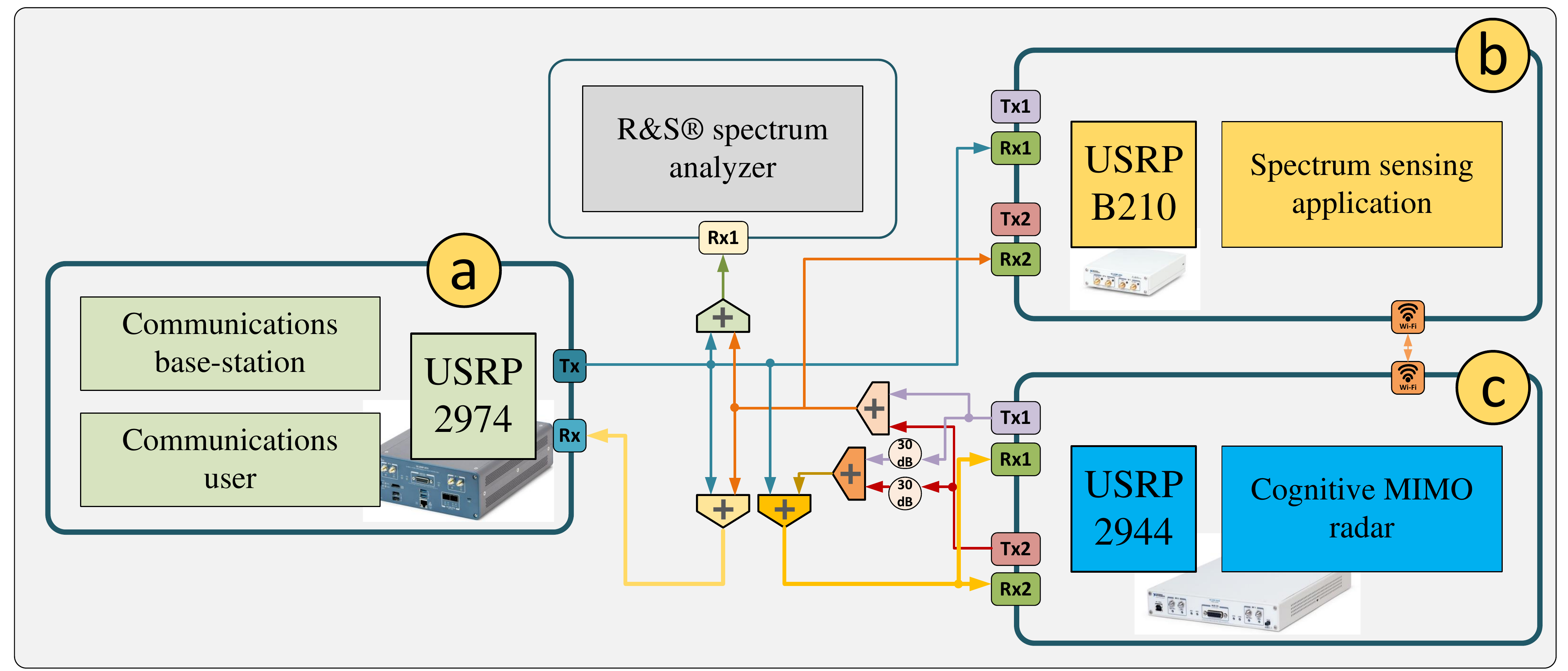}
    \caption{ Diagram depicting the connection of the JRC co-existence prototype \cite{Sensors}. 
    }
    \label{fig:ConnectionDiagram}
\end{figure}

In the sequel, a selection of experiments conducted using the developed prototype are presented and  the \gls{HW} results are analyzed. Given the frequency of operation and the bandwidths used, adequate separation is needed for target resolution. In the absence of a large experimental facility, for the practical applicability of the developed prototype  and  for the verification, a complete \gls{OTA} evaluation is discarded. Instead,  the connections shown in \figurename{~\ref{fig:ConnectionDiagram}} using \gls{RF} cables and splitters/ combiners are set-up, and the performance measured in a controlled environment.

\subsubsection{Target Generation}
\begin{figure}
    \centering
        \includegraphics[width=0.9\linewidth]{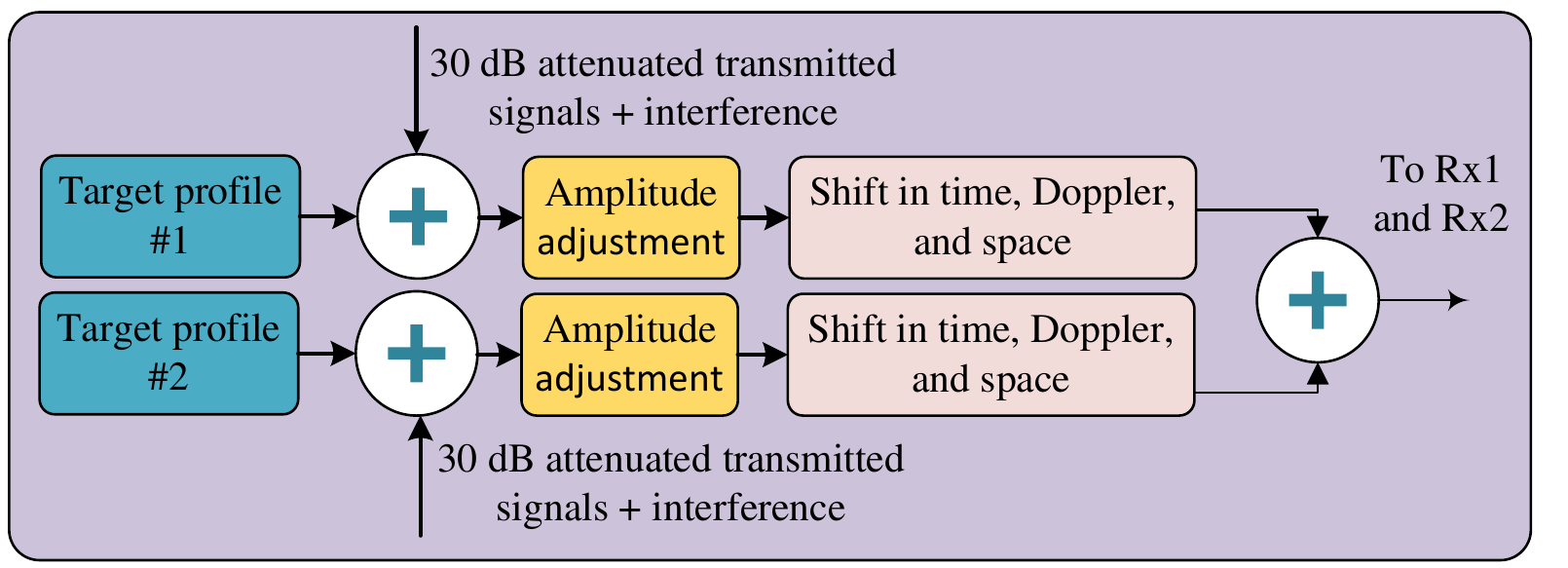}
    \caption{  Schematic depicting target generation  \cite{Sensors}.
    }
    \label{fig:targetGen}
\end{figure}

The transmit waveforms are attenuated using  $30$ dB attenuators  highlighted in \figurename{~\ref{fig:ConnectionDiagram}}. This attenuated signal is further shifted in time, frequency and spatial dimension in software to create a reflection from a target as depicted in  \figurename{~\ref{fig:targetGen}}. While a number of reflections, one corresponding to each target, can be generated, two targets are considered in the JRC prototype for ease of implementation. The target reflections are  further perturbed with the communications interference.  
%

%
\subsubsection{Transmit Waveforms}
Two choices are made available for the transmit waveforms: they could either be selected based on the options in \tablename{~\ref{tab:Apps}} or obtained from the adaptation in Section \ref{ssec:wave}.  During its execution, input parameters to optimize the waveforms provided by the \gls{GUI} to MATLAB, and the optimized set of sequences 
are returned to the application through the \gls{GUI}.  LabView G dataflow application is used to develop the remaining processing blocks of the radar system including matched filtering, Doppler processing, and scene generation.
Tables, \ref{tab:radar_param} and  \ref{tab:target_param} summarize the parameters  used for radar and targets in this experiment. 
\begin{table}[tbh]
\caption{Radar experiment parameter set-up \label{tab:radar_param}}{
\begin{tabular}{|c|c|}
\toprule
		\textbf{Parameters}                       & \textbf{Value}                     \\ 
		\hline
		Center frequency                           & $2$ GHz                    \\  \hline
		Real-time bandwidth        & $40$MHz \\ \hline
        Transmit and receive channels                           & $2 \times 2$                    \\ \hline
		Transmit power                      & $10$ dBm \\ \hline
		Transmit code length                      & $400$ \\ \hline
		Duty cycle                      & $50 \% $ \\  \hline
 		Range resolution                      & $3.7$ m \\
 		\hline
		Pulse repetition interval                      & $20 \mu $s \\  \hline
   \botrule
\end{tabular}}{\footnotetext[]}
\end{table}
\begin{table}[t!]
\caption{Radar target parameter set-up \label{tab:target_param}}{
\begin{tabular}{|c|c|c|}
\toprule
		\textbf{Parameters}                       & \textbf{Target 1} & \textbf{Target 2}                      \\ 
		\hline
		Range delay                     & $2 \mu$s   & $2.6 \mu$s  \\ \hline
		Normalized Doppler                       & $0.2$ Hz & $-0.25$ Hz\\ \hline
		Angle                     & $25 \deg$ &  $15 \deg$ \\ \hline
		Attenuation                       & $30$ dB & $35$ dB \\ \hline
 \botrule
\end{tabular}}{\footnotetext[]}
\end{table}
\subsubsection{Communication Set-up}
For the LTE communications, the downlink  between  a \gls{SISO} \gls{BS} and a single user with one antenna is established. It should however be noted that the experiments can be also be performed with uplink \gls{LTE}.  LabVIEW \gls{LTE} framework offers the possibility to vary the \gls{MCS} of \gls{PDSCH}, used for the transport of data between the \gls{BS} and the user, from $0$ to $28$ with the constellation size increasing from QPSK to $64$QAM \cite{LTE_NI}. \tablename{~\ref{tab:comm_param}} indicates  the  experimental  parameters for the communications. 
\begin{table}[t!]
\caption{Communication link parameter set-up \label{tab:comm_param}}{
\begin{tabular}{|c|c|}
\toprule 
		\textbf{Parameters}                       & \textbf{Value}                     \\ 
		\hline
		Communication \gls{MCS}         & \begin{tabular}[c]{@{}l@{}}MCS0 (QPSK $0.12$)\\ MCS10 (16QAM $0.33$) \\ MCS17 (64QAM $0.43$)\end{tabular} \\ \botrule
		Center frequency (Tx and Rx)          & $2$ GHz \\ \hline
		Bandwidth         & $20$ MHz \\
 \botrule
\end{tabular}}{\footnotetext[]}
\end{table}
\subsection{Performance Evaluation}
\subsubsection{Representative Results}
From Section \ref{ssec:wave} and following \cite{Sensors}, it is clear that when $\theta = 0$, the optimized waveforms are unable to exhibit notches at the undesired frequencies. The notches  appear gradually with an increase in $\theta$, with the deepest notch appearing when $\theta = 1$. However, when $\theta = 1$, the cross-correlation is  highest (phased-array), thereby diminishing the target resolution properties. The cross-correlation decreases with $\theta$, with the minimum being at $\theta = 0$. Thus an appropriate selection of $\theta$ leads to an optimized trade-off between spectral shaping and orthogonality.

Towards demonstrating the effect of an optimized radar waveform in a coexistence scenario,   $\theta = 0.75$ is selected  \cite{Sensors}.  Tables \ref{tab:target_param}  and \ref{tab:comm_param}, respectively, report the values are used for radar and communications. When transmitting a set of $M = 2$ waveforms, each of length $N = 400$, the radar occupies a bandwidth of $40$ MHz with  nulls to be obtained adaptively based on the  feedback from the spectrum sensing application. 

However, the \gls{LTE} communications framework has $20$ MHz bandwidth at its disposal for transmission. For full bandwidth transmission, the radar will not be able to perform satisfactorily due to the imposition of limited transmit power as a result of interference constraints. To simulate a meaningful spectrum co-existence scenario, some portions of this bandwidth are not used by the communications to enable their use by the radar. Towards this, the flexibility of the \gls{LTE} application is exploited and a resource block allocation of $1111111111110000000111111$ ($4$ physical resource blocks/bit) is considered. Here,  an entry ``$1$'' at a position indicates the use of the corresponding time-bandwidth resources in the \gls{LTE} application framework. 

\begin{figure*}[tbh]	
    \centering
    \subfigure[\gls{LTE} spectrum at R\&H spectrum analyzer.] {\label{fig:SpectrumMeasurment2a}\includegraphics[width=0.65\linewidth]{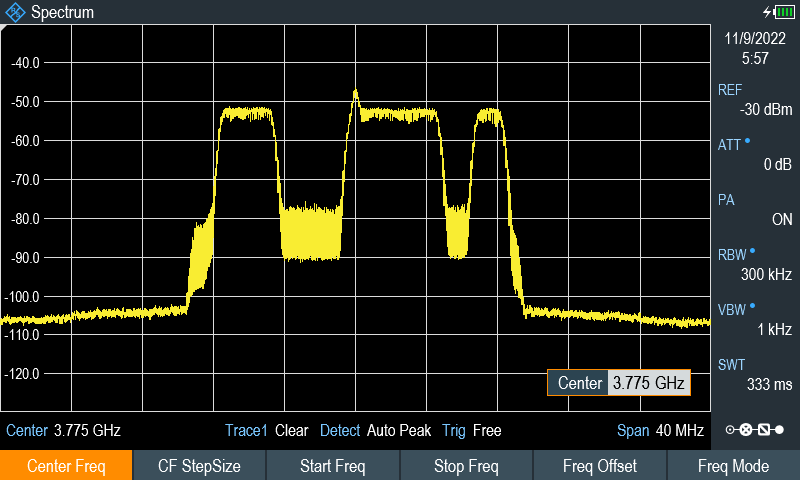}}
    \subfigure[Spectrum at the developed application.]{\label{fig:SpectrumMeasurment2b}\includegraphics[width=0.65\linewidth]{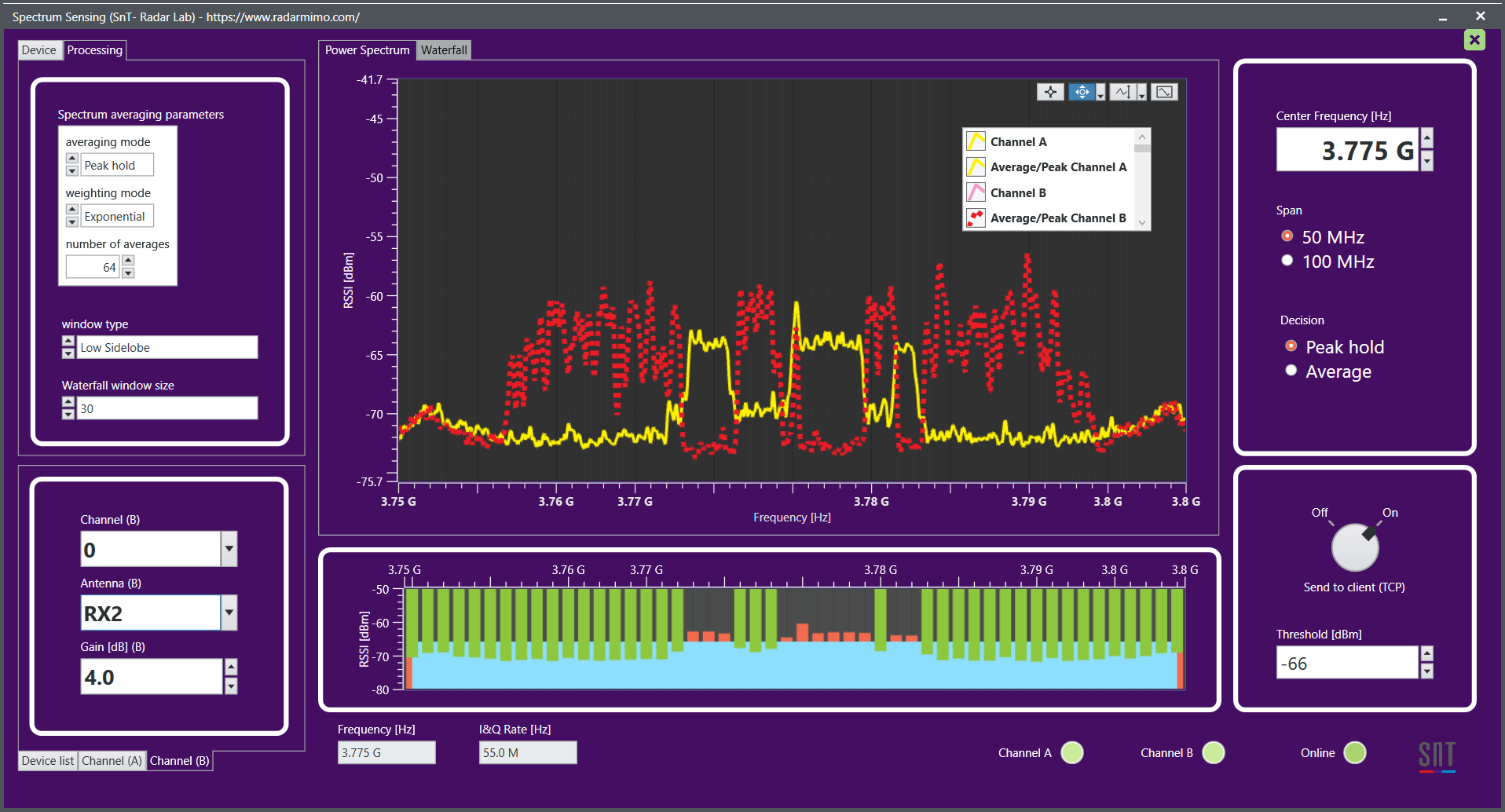}}    
\caption[]{ (a) Validation of the developed spectrum sensor using a commercial product for {LTE} downlink spectrum . 
(b)
Resulting spectrum of communications (yellow) and radar (dotted-red) at the developed two-channel spectrum sensing application \cite{Sensors}.}
\label{fig:SpectrumMeasurment2}
\end{figure*}

The spectrum of this \gls{LTE} downlink is measured with the developed spectrum sensing application as depicted in Figure \ref{fig:SpectrumMeasurment2}.  This figure serves two purposes, 
\begin{enumerate}
    \item Validates the developed spectrum analyzer application with a commercial product
    \item Indicates the achievement of the desired spectrum shaping objective    
\end{enumerate} 

\begin{figure*}[tbh]	
    \centering
    \subfigure[LTE in the presence of radar interference that occupies entire band by utilizing optimized sequences ($\theta = 0$).]{\label{fig:LTEandRadara}\includegraphics[width=\linewidth]{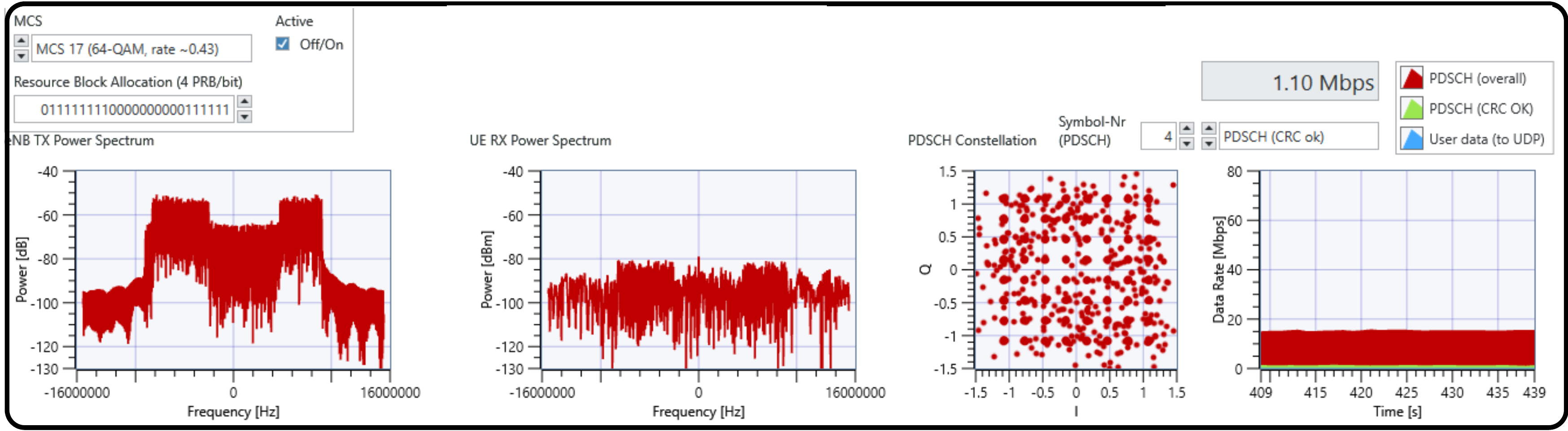}}
    \subfigure[LTE in the presence of optimized sequences for the coexistence scenario ($\theta = 0.75$).] {\label{fig:LTEandRadarb}\includegraphics[width=\linewidth]{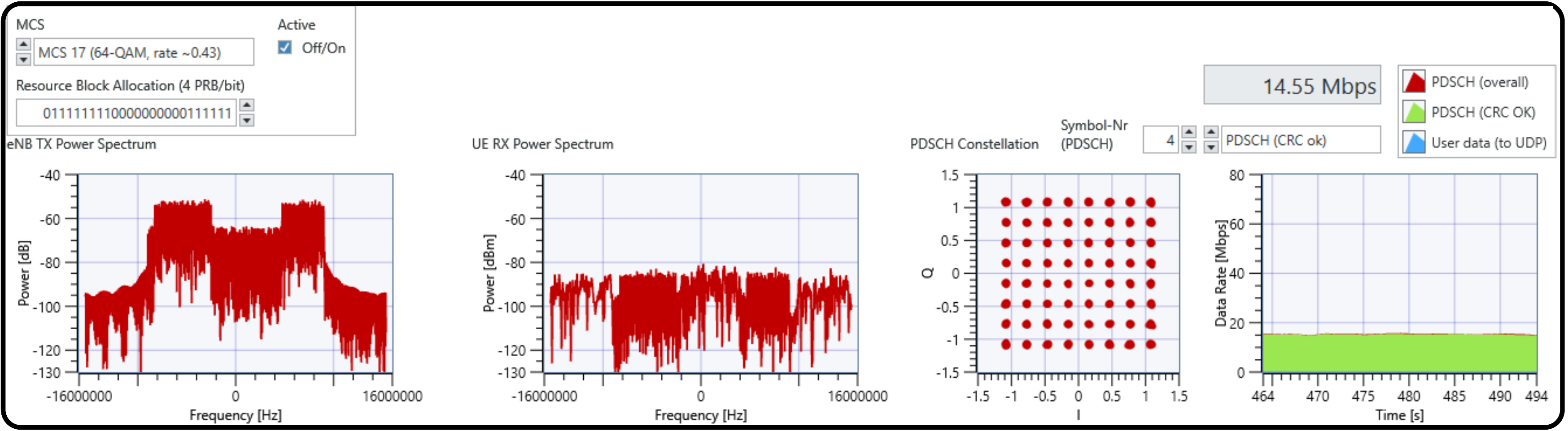}}
    \caption[]{(a), Radar is transmitting random-phase sequences on the communications frequency band  leading to a reduction in  throughput of the latter.  (b), Transmission of optimized MIMO radar waveforms enhances the performance of communications.}
    \label{fig:LTEandRadar1}
\end{figure*}
\begin{figure*}[tbh]	
    \centering
     \subfigure {\label{fig:LTEandRadarc}\includegraphics[width=0.8\linewidth]{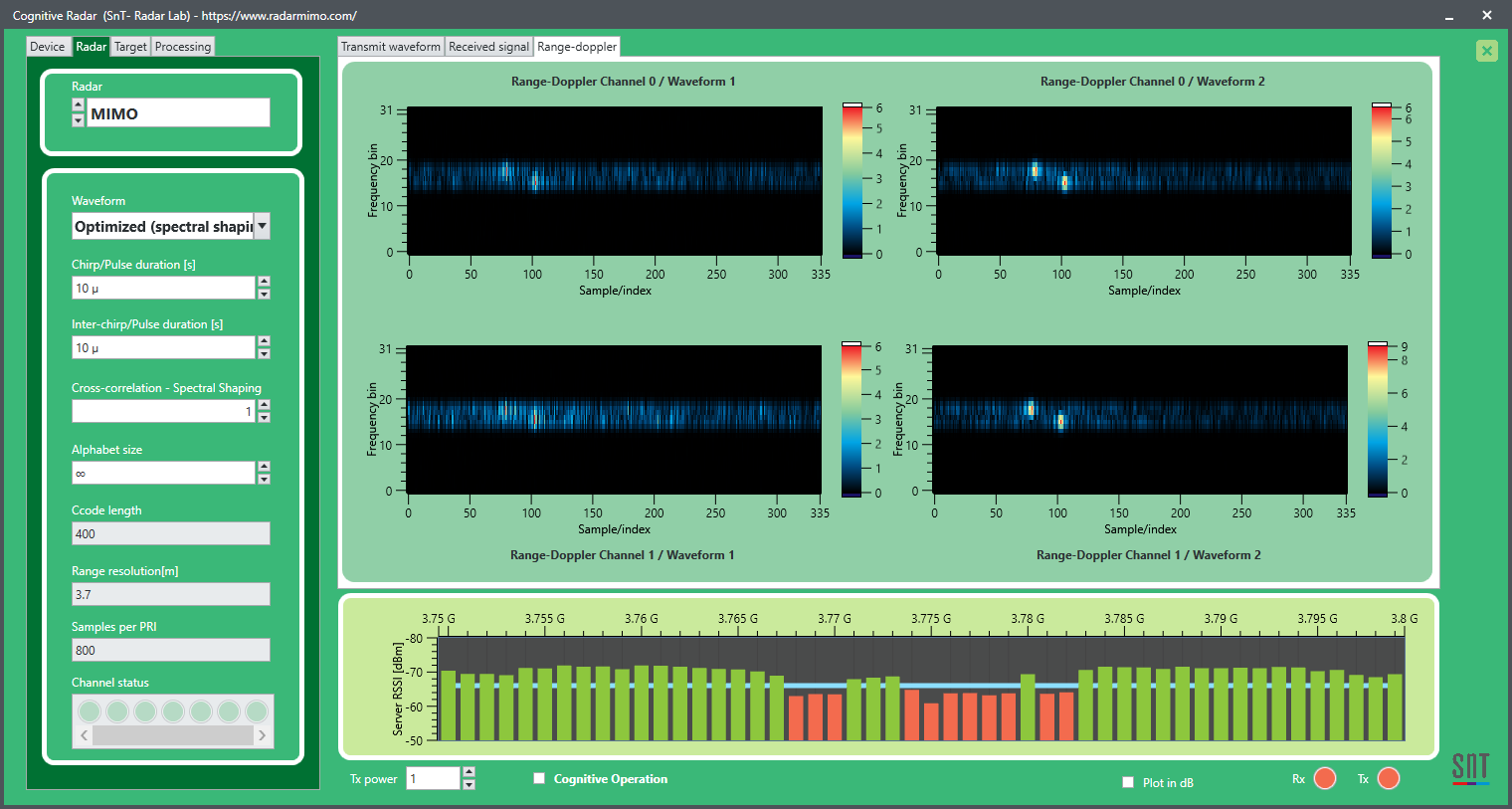}}

    \subfigure {\label{fig:LTEandRadard}\includegraphics[width=0.8\linewidth]{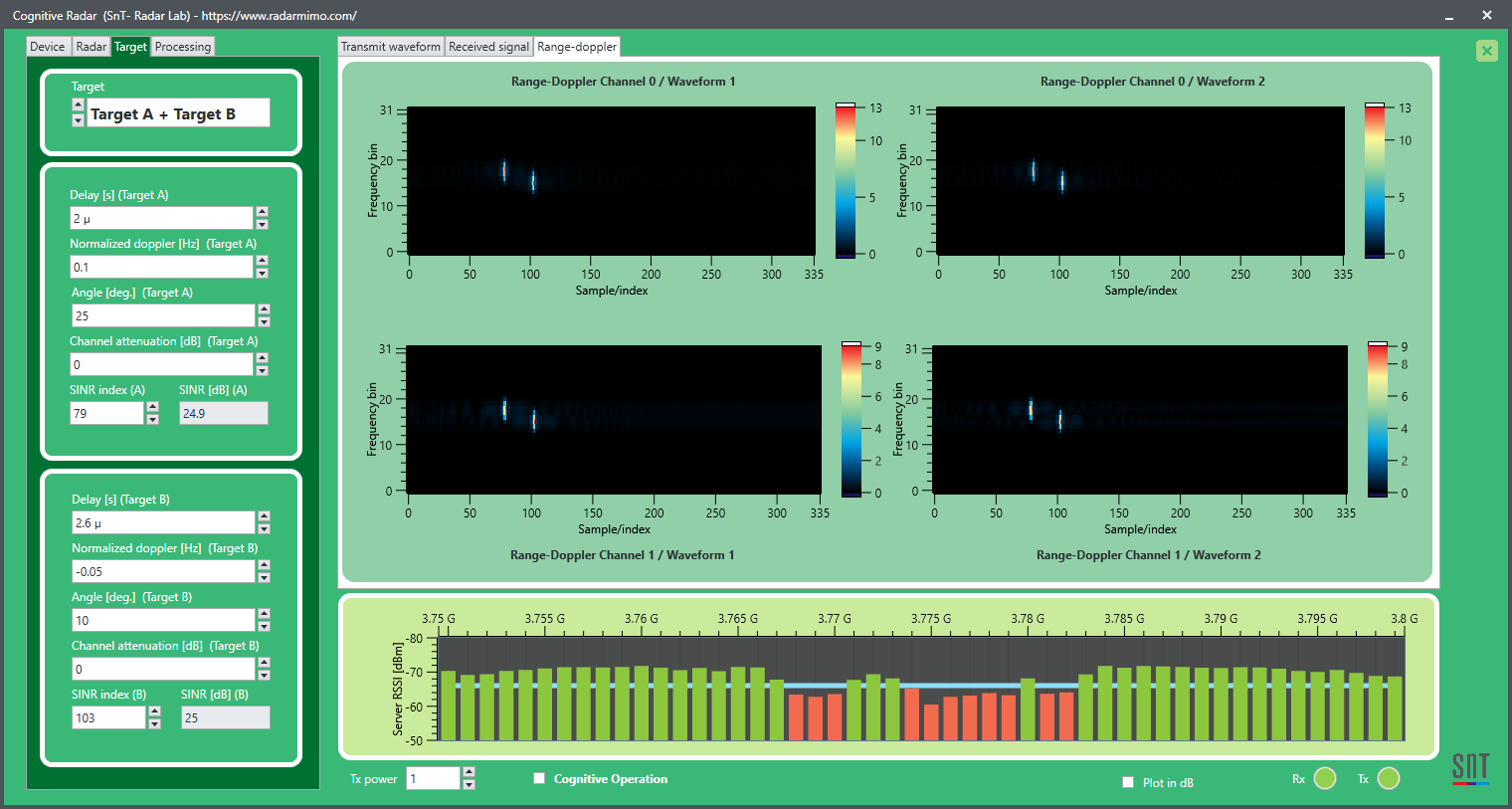}}
    \caption[]{ (a) MIMO radar utilizing optimized waveforms and not taking cognisance of communications transmissions ($\theta = 0$) results in interference from  communications which degrades the radar detection performance. (b) MIMO radar utilizing optimized sequences for enabling coexistence  ($\theta = 0.75$) resulting in the enhanced performance of radar.}
    \label{fig:LTEandRadar2}
\end{figure*}

When unaware of the spectrum use by the communications, the radar maximizes its target resolution capabilities by  transmitting optimized sequences with $\theta = 0$ and consuming the entire bandwidth. This leads to significant mutual interference, disrupting the  operations of both radar and communications, as shown in part (a) of figures \ref{fig:LTEandRadar1} and 
\ref{fig:LTEandRadar2}, thereby hampering their coexistence. In this situation, using the optimized waveforms obtained by $\theta = 0.75$ improves the performance of both systems, as illustrated in part (b) of figures  \ref{fig:LTEandRadar1} and \ref{fig:LTEandRadar2}.

\subsubsection{\gls{SINR} Performance Analysis}
\label{sssec:performance}
To further analyse the the performance of the proposed prototype,  the \gls{SINR} of the two targets is calculated for radar while  the \gls{PDSCH} throughput calculated by the \gls{LTE} application framework is reported. The experimentation methodology is highlighted below: 
\begin{enumerate}
	\item[Step-1: ] {\bf No Radar Transmission}: In the step,   the \gls{LTE} \gls{PDSCH} throughput for different MCS, i.e., MCS0, MCS10 and MCS 17, are collected. For each \gls{MCS},  $5, 10, 15$ and $20$ dBm of \gls{LTE} transmit power are used in accordance with link-budget.   
	\item[Step-2:]{\bf No \gls{LTE} Transmission}: In the stage,  the radar utilizes the entire bandwidth and generates optimized waveform by setting $\theta = 0$. The resulting received \gls{SNR} for the two targets is obtained numerically as the ratio of the peak power of the detected targets to the average power of the cells in the vicinity of the target location (on the range-Doppler map).
	\item[Step-3:] {\bf Concurrent Transmission}:A set of optimized radar waveforms for  $\theta = 0$ are transmitted along with the \gls{LTE} waveform, thereby causing significant mutual interference. The \gls{PDSCH} throughput as well as the \gls{SINR} of Target-$1$ and Target-$2$ are noted for MCS0, MCS10 and MCS17 and at $5, 10, 15$ and $20$ dBm power levels keeping the radar transmit power  fixed. To avoid bias in the results, the experiment is repeated $5$ times for each setting and the performance indicators are averaged.    
	\item[Step-4: ] {\bf Co-existence}: Step-3 is now repeated, but with the  waveforms optimized for $\theta = 0.75$ . 
\end{enumerate} 
 
\begin{figure*}[tbh]	
    \centering
    \subfigure{\label{fig:lte_mcs0}\includegraphics[width=90mm]{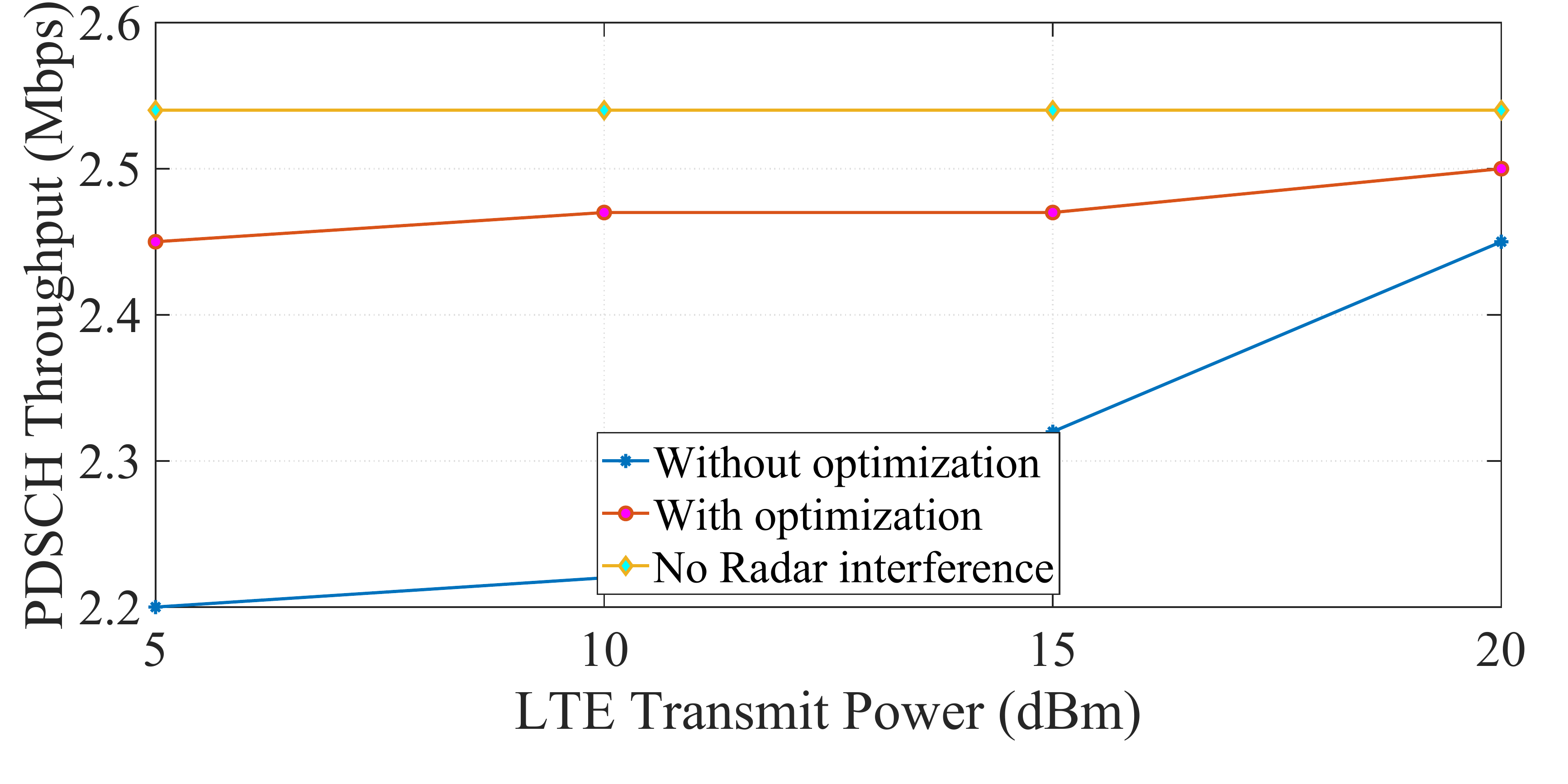}}
    \subfigure{\label{fig:lte_mcs10}\includegraphics[width=90mm]{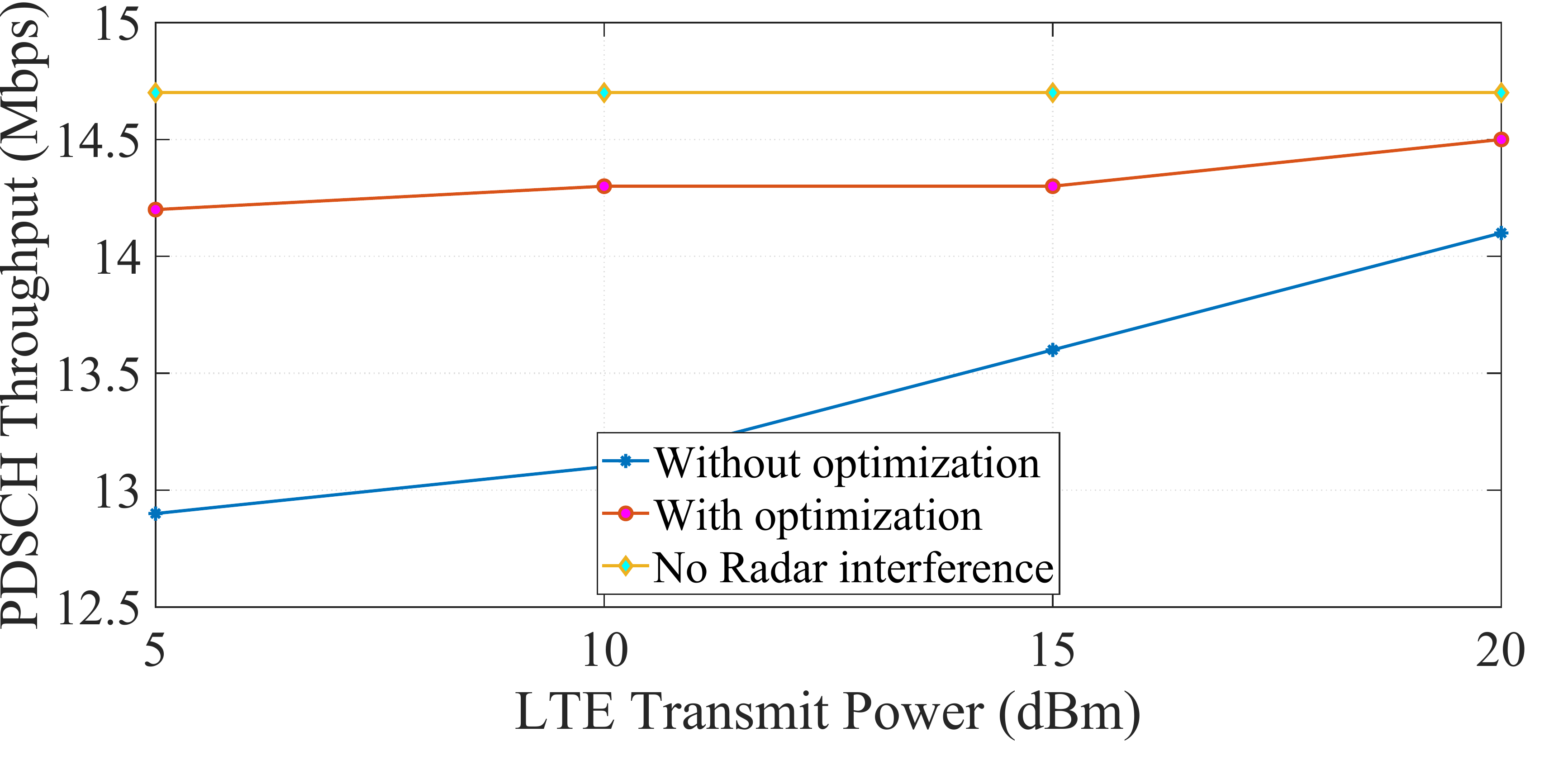}}
    \subfigure{\label{fig:lte_mcs17}\includegraphics[width=90mm]{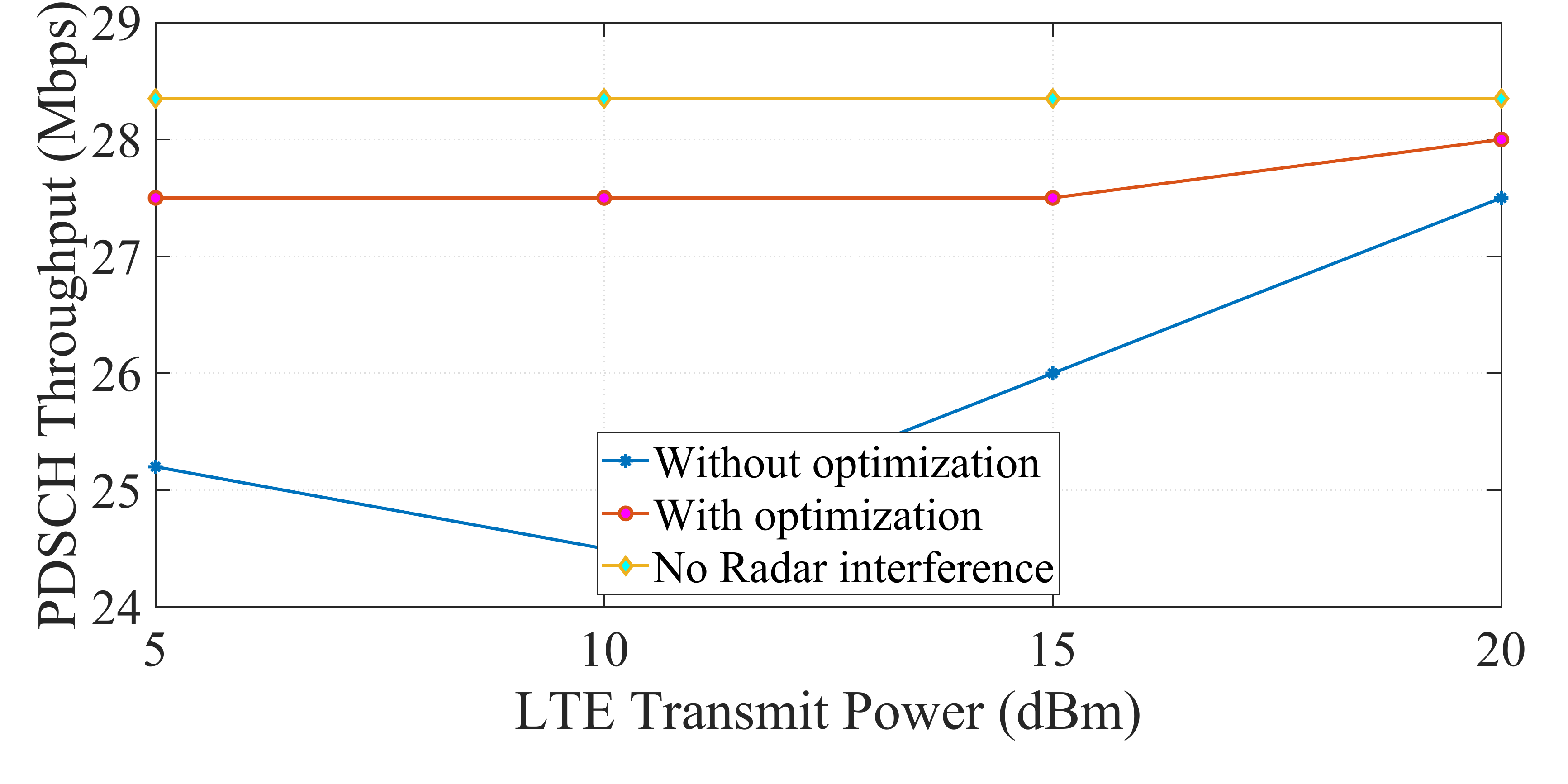}}
	\caption[]{PDSCH throughput of \gls{LTE} under radar interference for different MCS: (a) MCS 0 (QPSK $0.12$), (b) MCS 10 (16QAM $0.33$), (c) MCS 17 (64QAM $0.43$). Radar interference reduces \gls{PDSCH} throughput, but cognitive spectrum sensing followed by spectral shaping of the radar waveform improves \gls{PDSCH} throughput for all \gls{LTE} \gls{MCS}.
	\cite{Sensors}}\label{fig:lte-performance}
\end{figure*}

Figure \ref{fig:lte-performance} reports the obtained \gls{PDSCH} throughput, which is representative of the number of  successfully decoded bits \cite{Sensors}.  
This results shows that the link throughput degrades in the  presence of radar interference;  the degradation being higher for higher \gls{MCS}, since \gls{MCS}  are susceptible to variations in \gls{SINR}. Subsequently, the \gls{LTE} throughput improves when the radar optimizes its waveform with $\theta = 0.75$ and the improvement is prominent in the higher MCS.   
\begin{figure}[tbh]	
    \centering
    \subfigure{\label{fig:radar-performancelte_mcs0}\includegraphics[width=90mm]{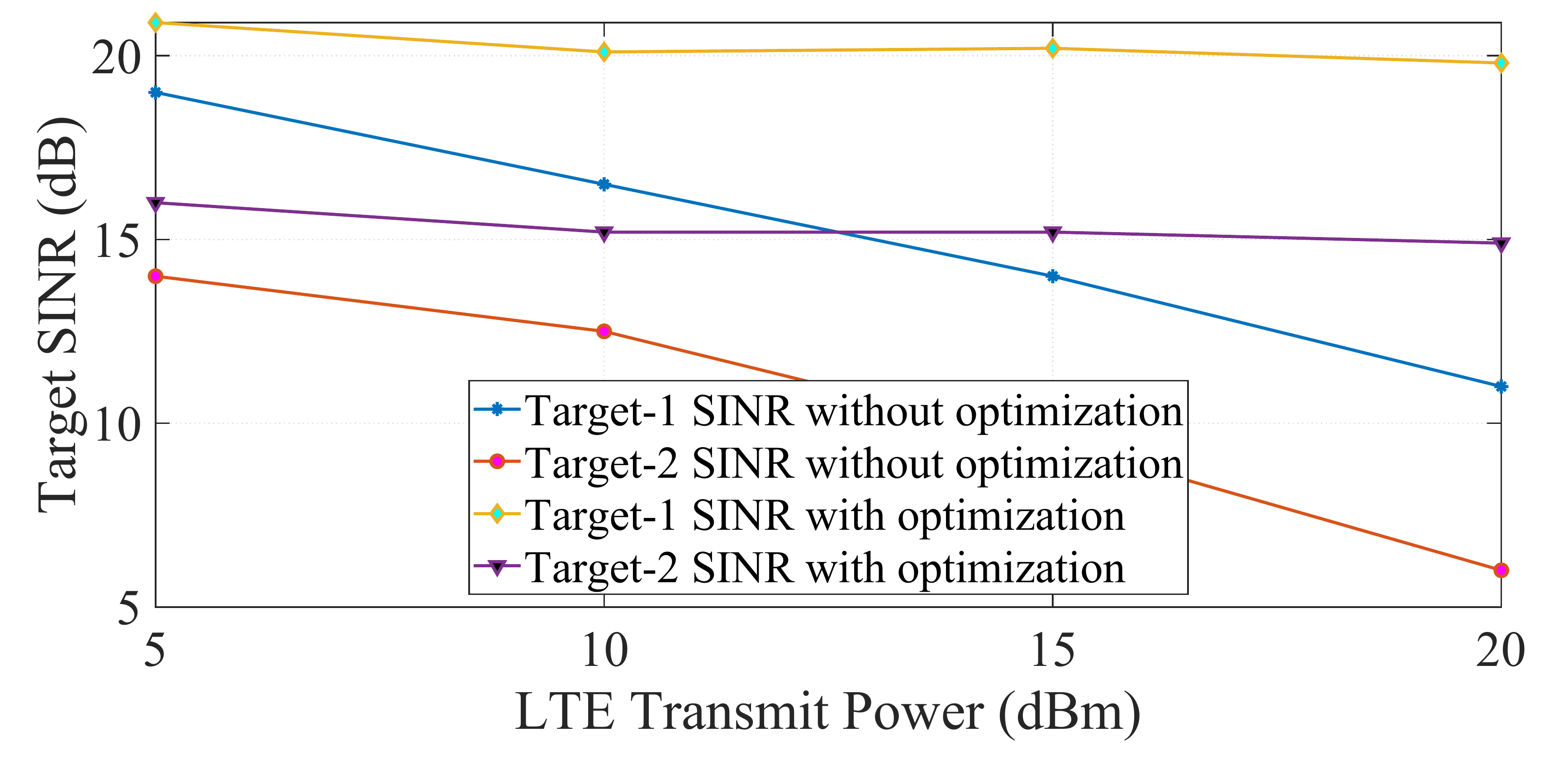}}
    \subfigure{\label{fig:radar-performancelte_mcs10}\includegraphics[width=90mm]{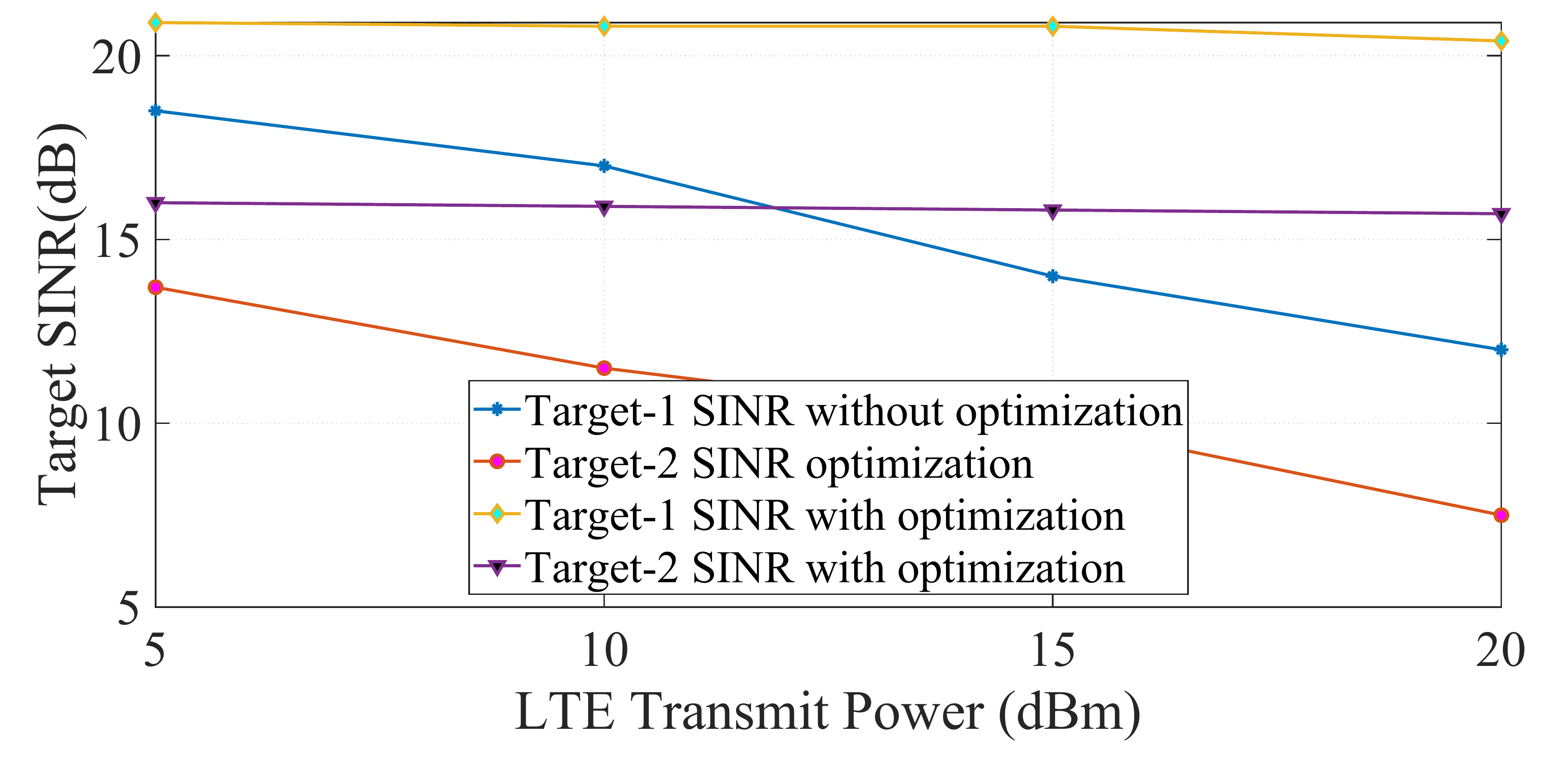}}
    \subfigure{\label{fig:radar-performancelte_mcs17}\includegraphics[width=90mm]{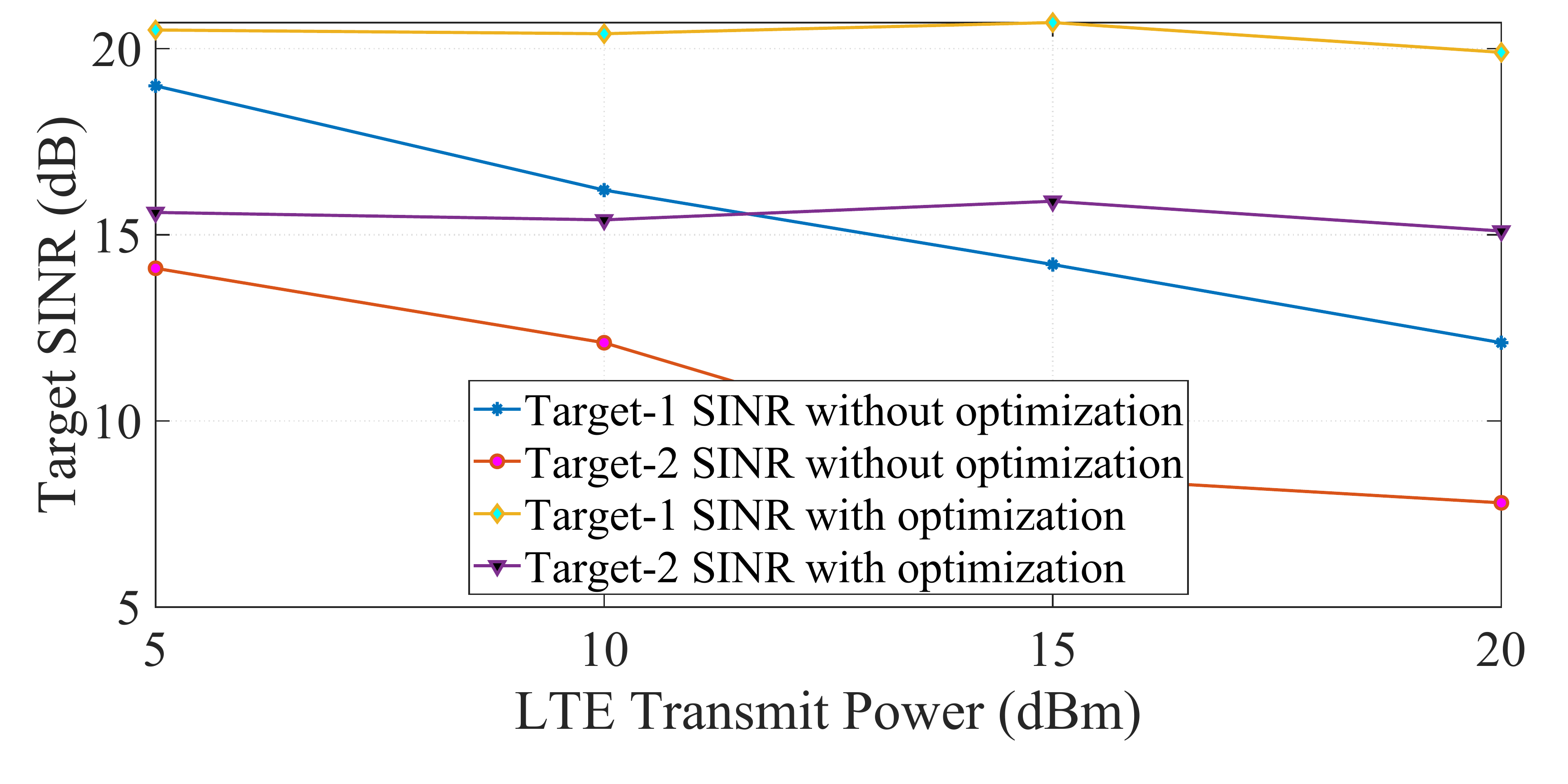}}

	\caption[]{Target \gls{SINR} under interference from downlink \gls{LTE} link for different MCS: (a) MCS 0 (QPSK $0.12$), (b) MCS 10 (16QAM $0.33$), (c) MCS 17 (64QAM $0.43$).forms are used. An upper bound on \gls{SNR} for the first and second targets  was $22$ dB and $17$ dB, respectively in the absence of communications interference.
	}\label{fig:radar-performance}
\end{figure}

Figure \ref{fig:radar-performance}, illustrates the  \gls{SINR} performance of the two targets  in the presence of \gls{LTE} interference.  \gls{SINR} improves for both targets when the radar optimizes the transmit waveforms by setting $\theta = 0.75$.
Interestingly, for high \gls{LTE} transmission power ($15$ dBm, and $20$ dBm), the  improvement  is higher; this results from the avoidance of the occupied \gls{LTE} bands. As a case in point, when the communications system is transmitting with a power of $20$ dBm, the use of  optimized waveforms enhances the target \gls{SINR}  in excess of $7$ dB over all the MCS values. Further,  the achieved \gls{SINR} of Target-1, and Target-2  in the absence of the \gls{LTE} interference serves as the benchmark. These values are is $22$ dB and $17$ dB respectively.  Finally, as a consequence of considering different attenuation paths for the two targets (kindly refer Table  \ref{tab:radar_param}), their measured \gls{SINR}s  are different.
\section{Other JRC prototypes}
\subsection{Low-rate co-existence prototype}
In \cite{cohen2017spectrum}, a spectral coexistence prototype based on Xampling (SpeCX) was mentioned. It employed a low-rate ADC which filters the received signal to predetermined frequencies before taking point-wise samples. These compressed samples, or ``Xamples", contain the information needed to recover the desired signal parameters. The SpeCX prototype consists of separate units of a cognitive radio receiver and an emulated cognitive radar transceiver. At the heart of the cognitive radio system lies a modulated wideband converter (MWC) that achieves the lower sampling rate bound and implements the sub-Nyquist analog front-end receiver. The card first splits the wideband signal into $4$ hardware channels, with an expansion factor of $5$, yielding $20$ virtual channels after digital expansion. In each channel, the signal is then mixed with a periodic sequence with $20\, \text{MHz}$ bandwidth that is generated on a dedicated FPGA. The sequences are chosen as truncated versions of Gold Codes \cite{gold1967optimal}, commonly used in telecommunication and satellite navigation. These were heuristically found to give good detection results in the MWC system, primarily due to small bounded cross-correlations within a set. This is useful when multiple devices are broadcasting in the same frequency range. 

Next, the modulated signal passes through an analog anti-aliasing low-pass filter (LPF). Specifically, a Chebyshev LPF of 7th order with a cut-off frequency ($-3\,\text{dB}$) of $50\,\text{MHz}$ was chosen for the implementation. Finally, the low rate analog signal is sampled by a National Instruments\textsuperscript{\textcopyright} ADC operating at sampling rate of $120\,\text{MHz}$ (with intended oversampling), leading to a total sampling rate of $480\,\text{MHz}$. The digital receiver is implemented on a National Instruments\textsuperscript{\textcopyright} PXIe-1065 computer with DC coupled ADC. Since the digital processing is performed at the low rate of $120 \, \text{MHz}$, very low computational load is required in order to achieve real time recovery. MATLAB\textsuperscript{\textregistered}and LabVIEW\textsuperscript{\textregistered} platforms are used for the various digital recovery operations. The sampling matrix is computed only once off-line using a calibration process.

The prototype is fed with radio-frequency (RF) signals composed of up to $5$ real comm transmissions, namely $10$ spectral bands with total bandwidth occupancy of up to $200\,\text{MHz}$ and varying support, with Nyquist rate of $6\,\text{GHz}$. 
Specifically, to test the system's support recovery capabilities, an RF input is generated using vector signal generators (VSG), each producing a modulated data channel with individual bandwidth of up to $20\,\text{MHz}$, and carrier frequencies ranging from $250\,\text{MHz}$ up to $3.1\,\text{GHz}$. The input transmissions then go through an RF combiner, resulting in a dynamic multiband input signal, that enables fast carrier switching for each of the bands. This input is specially designed to allow testing the system's ability to rapidly sense the input spectrum and adapt to changes, as required by modern cognitive radio and shared spectrum standards, e.g. in the SSPARC program. The system's effective sampling rate, equal to $480\,\text{MHz}$, is only $8\%$ of the Nyquist rate and 2.4 times the Landau rate. This rate constitutes a relatively small oversampling factor of $20\%$ with respect to the theoretical lower sampling bound. The main advantage of the Xampling framework, demonstrated here, is that sensing is performed in real-time from sub-Nyquist samples for the entire spectral range, which results in substantial savings in both computational and memory complexity. 

Support recovery is digitally performed on the low rate samples. The prototype successfully recovers the support of the comm transmitted bands. Once the support is recovered, the signal itself can be reconstructed from the sub-Nyquist samples in real-time. The reconstruction does not require interpolation to the Nyquist rate and the active transmissions are recovered at the low rate of $20\,\text{MHz}$, corresponding to the bandwidth of the aliased slices. The prototype's digital recovery stage is further expanded to support decoding of common comm modulations, including BPSK, QPSK, QAM and OFDM. There are no restrictions regarding the modulation type, bandwidth or other signal parameters, since the baseband information is exactly reconstructed regardless of its respective content.

By combining both spectrum sensing and signal reconstruction, the MWC prototype serves as two separate comm devices. The first is a state-of-the-art cognitive radio that performs real time spectrum sensing at sub-Nyquist rates, and the second is a unique receiver able to decode multiple data transmissions simultaneously, regardless of their carrier frequencies, while adapting to spectral changes in real time.

The cognitive radar system included a custom made sub-Nyquist radar receiver board composed of $4$ parallel channels which sample $4$ distinct bands of the radar signal spectral content. In each channel, the transmitted band with bandwidth $80 \, \text{KHz}$ is filtered, demodulated to baseband, and sampled at $250 \, \text{KHz}$ (with intentional oversampling). This way, four sets of consecutive Fourier coefficients are acquired. After sampling, the spectrum of each channel output is computed via fast Fourier transform and the 320 Fourier coefficients are used for digital recovery of the delay-Doppler map.
The prototype simulates transmission of $50$ pulses towards $9$ targets. The cognitive radar transmits over $4$ bands, selected according to an optimization procedure and occupying $3.2\%$ of the traditional wideband radar bandwidth.
\subsection{Index Modulation based co-design prototype}
\label{ssec:Index}
A particular prototype involving the co-design of the radar and communication functionality from a system level is presented in \cite{SpaCor}. the system considers different bands for the  radar and communication systems as well as separate waveforms; thus there is no interference amongst the two systems and the waveform selection is flexible. However, the use of multiband signalling leads to enhanced system complexity. Herein, the authors being in the concept of co-design and devise and demonstrate a  spatial modulation based communication-radar (SpaCoR) system. Central to the work is the use of  Index Modulation  through the  generalized spatial modulation (GSM) wherein the index of antenna element chosen is determined by the information bits to be transmitted; such a design, while reducing the number of active \gls{RF} chains, also induces spatial agility In this context, the authors in \cite{SpaCor} restrict an antenna element to transmit either the radar or communication waveform. To enhance the performance of such a system, an enhanced allocation of antennas to waveforms is presented in \cite{SpaCor} $-$ leading to a co-design. 

To demonstrate the feasibility of their system, the authors have implemented the {\em SpaCoR prototype} using a dedicated hardware. This versatile prototype features aspects enabling its reuse in different dual function systems:
\begin{table}[t!]
\caption{SpaCor Prototype Functionalities \label{tab:SpaCor}}{
\begin{tabular}{|c|c|}
\toprule 
	Baseband waveform generation  &  radar echo generation \\  
	\gls{OTA} signalling  & radar echo reception\\ \hline
	Frequency band waveform transmission & communication signal receptio \\ \hline
 \botrule
\end{tabular}}{\footnotetext[]}
\end{table}

The overall structure of the prototype combines the following
\begin{enumerate}
    \item Host Processing on a PC server offering a \gls{GUI} for setting the relevant parameter, generating the waveforms and processing the received signals. Once the paramerers are set, the JRC waveform is generated by the PC application at the onset of the experiment. Subsequently, the JRC waveform is transferred to the transmit board discussed next.
    \item A transmitter board comprising a field-programmable gate array (FPGA) board to realize a high speed data interface between the digital PC generated waveforms and the digital-to-analog convertor (DAC),  4DSP FMC216 DAC cards, and an up-conversion card for \gls{RF} transmission.
    \item  A 2D digital antenna array with 16 elements with the ability to control each element independently. The antenna works with carrier frequency $5.1$ GHz, has a bandwidth of $80$ MHz  and the size of each patch is $1.8$ cm $\times  1.3$ cm. The horizontal  and the vertical distance between two horizontally and the two vertically adjacent elements are both $2.7$ cm. The array is used in the configuration with 8 transmit and 8 receive antennas \cite{SpaCor}.
    \item A receiver board  converting the passband analog echoes and received waveforms to baseband digital streams for further processing on the host. The receiver board consists of a VC707 FPGA board, two FMC168 analog-to-digital convertors (ADCs) cards, and a radio frequency down-convertor board.
    \item A radar echo generator (REG) uses the transmitted waveform and generates the reflected echoes corresponding to  moving radar targets in an over-the-air setup. This unit consists of a Rhode \& Schwarz FSW signal and spectrum analyzer, which captures the received waveform, and a Rhode \& Schwarz SWM200 A vector signal generator, to add the set-up delays and Doppler shifts to the observed waveform for re-transmission. The transmit and receive  antennas are of dimensions $5$ cm $\times 5$ cm.
\end{enumerate}  
Experiments on the prototype are carried out with a radar pulse of $30~\mu~s$ communication baud rate of $0.4~MSps$ (Mega symbols per second), radar frequency band of $5.06-5.11$ GHz and the communication band of $5.11-5.14$ GHz. 16 bit DAC and ADCs are used with DAC update rate of $312.5~MSps$ and ADC sampling rate of $250MSps$. The results from \cite{SpaCor} indicates the  spatial agility induced by the GSM transmission over fixed antenna allocations. In particular, it improves the angular resolution and reduces the sidelobe level in the transmit beam pattern as well the Bit-error rate (BER) performance of the communication system.
\section{Conclusion}\label{Sec:Conclusion}
This chapter presents the need for JRC protoyping and discusses the requirements for combining existing communication or radar prototypes to enable joint functionality. A co-existence prototype is detailed and two other are summarized. These demonstrations brings the research in this emerging field already closer to the practitioners to enable early cross-fertilization of ideas and incorporation in upcoming initiatives like 6G. In addition, there have been a number of other important developments in the field of prototyping which is not captured in this book chapter. An useful initiative towards collating the requisite information in a single place is the Demonstration and Datasets Working Group (WG5) of the Integrated Sensing and Communications Emerging Technology Initiative from IEEE Communication Society ( https://isac.committees.comsoc.org/demonstrations-datasets/ ).

\bibliography{main}%

\latexprintindex

\end{document}